%% file: ms.tex
\documentclass[12pt,aps,rmp,showpacs]{revtex4}
\usepackage{amssymb,amsmath}

\newif\ifpdf
\ifx\pdfoutput\undefined
\pdffalse 
\else
\pdfoutput=1 
\pdftrue
\fi

\ifpdf
\usepackage[pdftex]{color}
\usepackage[pdftex]{graphicx}
\else
\usepackage{color}
\usepackage{graphicx}
\fi

\input macros	
\renewcommand{\.}{\ensuremath\M[0.075]}
\newcommand{\VP}[1]{\smash{\check#1\M[0.1]}}

\allowdisplaybreaks

\begin{document}

\ifpdf
\DeclareGraphicsExtensions{.pdf, .jpg, .tif}
\else
\DeclareGraphicsExtensions{.eps, .jpg}
\fi

\title{The Real Density Matrix}
\author{Timothy F. Havel}
\email[E-mail:\ ]{tfhavel@mit.edu}
\thanks{corresponding author.}
\affiliation{Dept.~of Nuclear Engineering,
Massachusetts Inst.~of Technology,
Cambridge, MA 02139}

\date{\today}

\begin{abstract}
We introduce a nonsymmetric real matrix which contains all the information that the usual Hermitian density matrix does, and which has exactly the same tensor product structure.
The properties of this matrix are analyzed in detail in the case of multi-qubit (e.g.~spin $= 1/2$) systems, where the transformation between the real and Hermitian density matrices is given explicitly as an operator sum, and used to convert the essential equations of the density matrix formalism into the real domain.
\end{abstract}

\pacs{03.65.Ca, 03.67.-a, 33.25.+k, 02.10.Xm}

\maketitle

\vspace{-0.75cm}
\section{Prologue}
The density matrix plays a central role in the modern theory of quantum mechanics, and an equally important role in its applications to optics, spectroscopy, and condensed matter physics.
Viewed abstractly, it is a self-adjoint operator $\rho$ on system's Hilbert space, the expectation values $0 \le \BRA{\psi}\,\rho\,\KET{\psi} \le 1$ of which give the probability of observing the system in the state $\KET{\psi}$.
As a matrix, however, it is generally represented versus the operator (or ``Liouville'') basis $\KET{i}\BRA{j}$ induced by a choice of a complete orthonormal basis $\{\KET{i}\mid i=0,1,\ldots\}$ in the underlying Hilbert space.
These complex-valued matrices $\RHO \equiv [\BRA{i}\rho\KET{j}]_{i,j}$ are necessarily Hermitian and positive semi-definite.
Their diagonal entries are the probabilities of these mutually exclusive basis states, whereas their off-diagonal entries prescribe the amounts by which the probabilities of their coherent superpositions deviate from the corresponding classical mixtures due to interference.

Another option is to use a operator basis the elements of which have rank exceeding one, so that it is not induced by any Hilbert space basis.
The most common example here is the representation of operators on a two-dimensional Hilbert space by real linear combinations of Pauli matrices $\{ \SIG[0] (\equiv \MAT I_{2\LAB D}), \SIG[1], \SIG[2], \SIG[3] \}$.
In this case the basis elements themselves are self-adjoint and so can be given a physical interpretation, e.g.~as the components of the Bloch or Stokes vector \cite{Bloch:46,FeyVerHel:57}.
Although arbitrary bases of self-adjoint operators could be used, for multi-particle systems it is desirable that the overall basis be induced by identical bases on each particle's Hilbert subspace.
With the Pauli matrices this leads to the so-called \emph{product operator} representation \cite{ErnBodWok:87}, in which density operators are represented by linear combinations of all possible tensor (\emph{ergo} Kronecker) products of the Pauli matrices, e.g.~$\SIG[k]^1\SIG[\ell]^2 \equiv (\SIG[k] \otimes \SIG[0]) (\SIG[0] \otimes \SIG[\ell])$.
In contrast to the Hermitian case, it has not been widely recognized that this tensor product structure is reflected by the real-valued coefficients in the expansion of any density operator in terms of product operators (also known as the \emph{coherence vector} \cite{MahleWeber:98}).
Thus if one properly arranges these coefficients in a matrix one obtains a \emph{real} but \emph{nonsymmetric} analog of the Hermitian density matrix with the \emph{same} tensor product structure.
This fact holds for any number of Hilbert spaces $\mathfrak H_k$ ($k = 1,2,\ldots$) of arbitrary (even infinite) dimension $L > 0$ and self-adjoint bases $\{ B^k_\ell \}_{\ell=1}^L$ for the space of bounded linear operators on each.

The purpose of this paper is show how, in the case of multi-qubit (\emph{ergo} two-state quantum) systems, one can perform all the usual Hermitian density matrix calculations entirely with the real density matrix.
While the formulae are more complicated in most cases than they are with the Hermitian density matrix, we argue that they are in many respects closer to the underlying physics than the Hermitian formulae, simply because the entries of the real density matrix correspond to (expectation values of) observables.
Indeed, it is well-known that the single-qubit Pauli algebra is nothing but a complex matrix representation of the geometric (or Clifford) algebra of a three-dimensional Euclidean vector space \cite{HavelDoran:02}.
This \emph{real} algebra in turn has been demonstrated to be a concise but versatile formalism within which to analyze and teach much of modern physics \cite{Hestenes:03,DoranLasen:03}.
It makes a certain amount of sense to use a representation wherein the ``reverse-even'' entities in the algebra, i.e.~scalars and vectors, are real while the less-familiar ``reverse-odd'' entities, i.e.~``pseudo-scalars'' and ``pseudo-vectors'', are purely imaginary (cf.~\citet{Baylis:99}).
The drawback of our representation is that the matrices no longer form a representation of the underlying geometric algebra, i.e.~the geometric product no longer corresponds to matrix multiplication.
We will leave it to the community to decide if or when the advantages outweigh the disadvantages, and offer our results simply as the outcome of an intellectual exercise.

\section{Metamorphosis} \label{sec:meta}
We begin by introducing a bit of notation which will considerably simplify the remainder of our presentation.
First, instead of the above bra-ket notation, let us write the $2\times2$ elementary matrices as
\begin{equation}
\MAT E_{00} \leftrightarrow \KET0\BRA0 \,,\M \MAT E_{10} \leftrightarrow \KET1\BRA0 \,,\M \MAT E_{01} \leftrightarrow \KET0\BRA1 \,,\M \MAT E_{11} \leftrightarrow \KET1\BRA1 \,,
\end{equation}
where $\KET0 \leftrightarrow \MAT e_0$, $\KET1 \leftrightarrow \MAT e_1$ denote an orthonormal basis for a two-dimensional Hilbert space.
Then it is easily seen that, for any nonnegative integers $i, j \le M \equiv 2^N - 1$, the $(M+1)\times(M+1)$ elementary matrix $\MAT E_{ij}$ is the Kronecker product of $2\times2$ elementary matrices the indices of which are the bits $i_n, j_n \in \{ 0,\,1 \}$ in the binary expansions of $i, j$, respectively, i.e.
\begin{equation}
\MAT E_{ij} ~\equiv~ \big[ \delta_{ik} \delta_{j\ell} \big]_{k,\ell=0}^{M,M} ~=~ \MAT E_{i_1j_1} \otimes\cdots\otimes \MAT E_{i_Nj_N} ~,
\end{equation}
where the $\delta$'s are Kronecker deltas. 
In an analogous fashion, we will denote the usual $2\times2$ Pauli matrices by
\begin{equation}
\MAT P_{00} \M[0.5]\equiv\M[0.5] \SIG[\,0] \,,\M \MAT P_{10} \M[0.5]\equiv\M[0.5] \SIG[\,1] \,,\M \MAT P_{01} \M[0.5]\equiv\M[0.5] \SIG[\,2] \,,\M \MAT P_{11} \M[0.5]\equiv\M[0.5] \SIG[\,3] \,.
\end{equation}
In this notation it may readily be verified that the multiplication table among the Pauli matrices may be expressed succinctly as
\begin{equation}
\MAT P_{ij\,} \MAT P_{k\ell} ~=~ \imath^{\,(i\ell-jk)(1-2ij)(1-2k\ell)}\, \MAT P_{(i+k-2ik),\M[0.05](j+\ell-2j\ell)} \quad\big( i,j,k,\ell \in \{0,\,1\} \big) ~,
\end{equation}
where $\imath^2 = -1$.
This indexing scheme may be extended to all Kronecker products of these matrices in the same way as for the $2\times2$ elementary matrices, i.e.
\begin{equation}
\MAT P_{ij} ~=~ \MAT P_{i_1j_1} \otimes\cdots\otimes \MAT P_{i_Nj_N} \quad(0 \le i,j \le M) ~.
\end{equation}
For example, if $M = 3$ we have $\MAT P_{01} = \SIG[0]\otimes\SIG[2]$, $\MAT P_{02} = \SIG[2]\otimes\SIG[0]$, and $\MAT P_{03} = \SIG[2]\otimes\SIG[2]\,$.

The Hermitian density matrix, of course, can be expanded relative to either the elementary matrix basis or the Pauli matrix basis, e.g.
\begin{equation}
\RHO ~=~ {\sum}_{i,j=0}^{\,1,\,1}\; \rho_{ij}\, \MAT E_{ij} ~=~ \text{\large$\tfrac12$} ~ {\sum}_{i,j=0}^{1,\,1}\; \sigma_{ij}\, \MAT P_{ij} ~, \label{eq:expansion}
\end{equation}
for a single qubit with $\rho_{ij} \in \FLD C$ but $\sigma_{ij} \in \FLD R$.
Both of these bases are orthogonal relative to the Hilbert-Schmidt inner product, which is given by $\HIP{\MAT X}{\MAT Y} \equiv \TR(\MAT{XY}^\dag) \equiv 2^N\, \langle\, \MAT{XY}^{\dag \,} \rangle$ for any number of qubits $N > 0$.
Thus there is a unique unitary superoperator $2^{\.-1/2}\, \ALG U$ which carries the Pauli to the elementary matrix basis, where the factor of $\sqrt2$ comes from $\| \MAT P_{ij} \|^2 \equiv \HIP{\MAT P_{ij}}{\MAT P_{ij}} = 2\, \| \MAT E_{ij} \|^2$.
This superoperator, moreover, is nearly self-adjoint since $\HIP{\MAT E_{ij}}{\MAT P_{k\ell}} = \HIP{\MAT E_{k\ell}}{\MAT P_{ij}}$ for all $0\le i,j,k,\ell \le 1$ with the sole exception of $\HIP{\MAT E_{10}}{\MAT P_{01}} = \imath\, \HIP{\MAT E_{01}}{\MAT P_{10}}$ and its complex conjugate.
On applying $\ALG U$ to both sides of Eq.~(\ref{eq:expansion}), therefore, we obtain
\begin{equation}
\ALG U(\RHO) ~=~ \text{\large$\tfrac12$} ~ {\sum}_{i,j=0}^{\,1,\,1}\; \rho_{ij}\. \sqrt{{\imath}^{\,j\.-\.i}/2^{\.|j-i\.|}}\, \big(\. \MAT P_{ji} \:+\: {(-1)}^{(1-\.i\.)\.j}\, \MAT P_{ij\.} \big) ~=~ {\sum}_{i,j=0}^{1,\,1}\; \sigma_{ij}\, \MAT E_{ij} ~.
\end{equation}
We will take this as our definition of the \emph{real density matrix} for a single qubit.
Henceforth, we shall denote this by
\begin{equation}
\begin{bmatrix} ~\sigma_{00}~&~\sigma_{01}~ \\ ~\sigma_{10}~&~\sigma_{11}~ \end{bmatrix} \M\equiv\M \SIG \M\equiv\M \ALG U(\RHO) \M=\M 2 \begin{bmatrix} \AVG{\RHO\, \SIG[0]} & \AVG{\RHO\, \SIG[2]} \\ \AVG{\RHO\, \SIG[1]} & \AVG{\RHO\, \SIG[3]} \end{bmatrix} ~.
\end{equation}
Note that our choice of normalization gives $\sigma_{00} = 2\. \langle\, \RHO\, \rangle = 1$, so that although $\ALG U$ is otherwise unitary the Hilbert-Schmidt norm is scaled by a factor of $\sqrt2$; specifically $2\, \|\RHO\M[0.1]\|^2 =\|\SIG\M[0.15]\|^2 \equiv 2\. \langle \SIG^{\!\top\!} \SIG\M[0.1] \rangle = 1 + \sigma_{10}^2 + \sigma_{01}^2 + \sigma_{11}^2$.

Let us now look at some explicit representations of the superoperator $\ALG U$.
To begin with, the mapping from the Pauli to the basis elementary is clearly
\begin{equation}
\begin{aligned} \MAT E_{00} ~=\M[0.75] & \HALF\big( \MAT P_{00} + \MAT P_{11} \big) \\
\MAT E_{10} ~=\M[0.75] & \HALF\big( \MAT P_{10} - \imath\, \MAT P_{01} \big) \end{aligned} \qquad
\begin{aligned} \MAT E_{01} ~=\M[0.75] & \HALF\big( \MAT P_{10} + \imath\, \MAT P_{01} \big) \\
\MAT E_{11} ~=\M[0.75] & \HALF\big( \MAT P_{00} - \MAT P_{11} \big) ~, \end{aligned}
\end{equation}
and hence (since coordinates are contravariant)
\begin{equation} \label{eq:usvd}
\KET{\SIG} ~\equiv~ \begin{bmatrix} 1\\ \sigma_{10}\\ \sigma_{01}\\ \sigma_{11} \end{bmatrix} \M[0.4]=\M[0.4] \begin{bmatrix} ~1~&0&0&0~\\ ~0~&1&0&0~\\ ~0~&0&-\imath&0~\\ ~0~&0&0&1~ \end{bmatrix} \M[-0.5] \begin{bmatrix} ~1&~0&0&1\\ ~0&~1&1&0\\ ~0&~1&-1&0\\ ~1&~0&0&\!-1 \end{bmatrix} \M[-0.5] \begin{bmatrix} \rho_{00}\\ \rho_{10}\\ \rho_{01}\\ \rho_{11} \end{bmatrix} \M[0.5]\equiv\M[0.6] \EMB{\ALG{VW}}\, \KET{\RHO} \M[0.5]\equiv\M[0.6] \EMB{\ALG U}\, \KET{\RHO} ~,
\end{equation}
where we have factored the overall superoperator's matrix $\EMB{\ALG U}$ into the product of a diagonal matrix $\EMB{\ALG V}$ and a purely real one $\EMB{\ALG W}$.
An operator sum representation for the superoperator $\ALG W$ may be derived from the singular value decomposition of its \emph{Choi matrix}, i.e.
\begin{equation}
\begin{bmatrix} ~1&~0&~0&\!-1\\ ~0&~1&~1&0\\ ~0&~1&~1&0\\ ~1&~0&~0&\!-1 \end{bmatrix} ~=~ \begin{bmatrix} ~1&~0&\!-1&0\\ ~0&~1&0&1\\ ~0&~1&0&\!-1\\ ~1&~0&1&0 \end{bmatrix} \begin{bmatrix} ~1\:&\:0\:&\:0\:&\:0~\\ ~0\:&\:1\:&\:0\:&\:0~\\ ~0\:&\:0\:&\:0\:&\:0~\\ ~0\:&\:0\:&\:0\:&\:0~ \end{bmatrix} \begin{bmatrix} 1&0&~1&0\\ 0&1&~0&1\\ 0&1&~0&\!-1\\ \!-1&0&~1&0 \end{bmatrix}^{\displaystyle\top} ,
\end{equation}
where the Choi matrix (left) is obtained simply by swapping certain pairs of the entries in $\EMB{\ALG W} \equiv \big[ w_{ij} \big]_{i,j=0}^{\,3,\,3}$ \cite{Havel!QPT:03}; specifically $w_{20} \leftrightarrow w_{01}$, $w_{30} \leftrightarrow w_{11}$, $w_{22} \leftrightarrow w_{03}$ and $w_{32} \leftrightarrow w_{13}$.
Observe that the left-singular vectors associated with the nonzero singular values of $\EMB{\ALG W}$ can be written as ``columnized'' Pauli matrices, specifically $\KET{\MAT P_{00}}$ and $\KET{\MAT P_{10}}$ in the notation of Eq.~(\ref{eq:usvd}), while the corresponding right-singular vectors are $\KET{\MAT P_{11}}$ and $\KET{\MAT P_{10}\,}$.
It follows that the operator sum form of $\ALG W$ is \cite[Proposition~3]{Havel!QPT:03}
\begin{equation}
\ALG W(\RHO) ~=~ \MAT P_{00}\, \RHO\, \MAT P_{11} \,+\, \MAT P_{10}\, \RHO\, \MAT P_{10} ~=~ \begin{bmatrix} \rho_{00} + \rho_{11} & \rho_{10} - \rho_{01} \\ \rho_{10} + \rho_{01} & \rho_{00} - \rho_{11} \end{bmatrix} ~.
\end{equation}

Although an operator sum form for $\ALG V$ could be obtained by this same approach, since it is diagonal a more compact representation of its action may be obtained by packing its nonzero entries into a single $2\times2$ matrix which acts via the ``entrywise'' or \emph{Hadamard product} ``$\odot$'' \cite{HaShViCo:01}, as follows:
\begin{equation} \begin{split}
\SIG ~=~ \MAT Q \odot \big( \RHO\, \MAT P_{11} \,+\, \MAT P_{10}\, \RHO\, \MAT P_{10} \big) ~\equiv\M & \begin{bmatrix} ~1&-\imath~\\ ~1&~1~ \end{bmatrix} \odot \begin{bmatrix} \rho_{00} + \rho_{11} & \rho_{10} - \rho_{01} \\ \rho_{10} + \rho_{01} & \rho_{00} - \rho_{11} \end{bmatrix} \\
\equiv\M& \begin{bmatrix} ~\rho_{00} + \rho_{11} ~&~ \imath\, (\rho_{01} - \rho_{10})~ \\ ~\rho_{01} + \rho_{10} ~&~ \rho_{00} - \rho_{11}~ \end{bmatrix} ~.
\end{split} \end{equation}
Since $\ALG W$ is self-adjoint and the overall superoperator $\ALG U$ is unitary (up to a factor of $\sqrt2$), it is easily seen that the inverse $\ALG U^{-1}$ can be written as
\begin{equation}
\RHO ~=~ \HALF \left( \big(\M[0.05] \OL{\MAT Q} \odot \SIG \big)\, \MAT P_{11} ~+~ \MAT P_{10}\, \big(\M[0.05] \OL{\MAT Q} \odot \SIG \big)\, \MAT P_{10}\, \right) ~,
\end{equation}
where the overbar indicates the complex conjugate of all the matrix entries.

The beauty of this operator sum form for the superoperator $\ALG U$ is that the Hadamard product obeys the mixed product formula with the Kronecker product,
\begin{equation} \label{eq:mixed}
(\MAT A \otimes \MAT B) \odot (\MAT C \otimes \MAT D) ~=~ (\MAT A \odot \MAT C) \otimes (\MAT B \odot \MAT D) ~,
\end{equation}
just like the usual matrix product does.
Thus if we extend $\ALG U$ to factorable multi-qubit density matrices in the obvious way,
\begin{equation}
\ALG U(\RHO^1 \otimes\cdots\otimes \RHO^N) ~\equiv~ \ALG U(\RHO^1) \otimes\cdots\otimes \ALG U(\RHO^N) ~,
\end{equation}
and thence to arbitrary multi-qubit density matrices by linearity, we immediately obtain a general expression.
Explicitly, in the case of two qubits, we get
\begin{equation} \begin{split}
\SIG \M\equiv\M & \SIG^1 \otimes \SIG^2 \M\equiv\M \ALG U(\RHO^1) \otimes \ALG U(\RHO^2) \\
=\M & \Big( \MAT Q \odot \big( \RHO^1\, \MAT P_{11} + \MAT P_{10}\, \RHO^1\, \MAT P_{10} \big)\! \Big) \otimes \Big( \MAT Q \odot \big( \RHO^2\, \MAT P_{11} + \MAT P_{10}\, \RHO^2\, \MAT P_{10} \big)\! \Big) \\
=\M & \big( \MAT Q \otimes \MAT Q \big) \odot \Big(\! \big( \RHO^1\, \MAT P_{11} + \MAT P_{10}\, \RHO^1\, \MAT P_{10} \big) \otimes \big( \RHO^2\, \MAT P_{11} + \MAT P_{10}\, \RHO^2\, \MAT P_{10} \big)\! \Big) \\
=\M & \big( \MAT Q \otimes \MAT Q \big) \odot \big( \begin{aligned}[t] & \!
(\RHO^1\, \MAT P_{11}) \otimes (\RHO^2\, \MAT P_{11}) + (\RHO^1\, \MAT P_{11}) \otimes (\MAT P_{10}\, \RHO^2\, \MAT P_{10}) ~\cdots \\
& +\, (\MAT P_{10}\, \RHO^1\, \MAT P_{10}) \otimes (\RHO^2\, \MAT P_{11}) + (\MAT P_{10}\, \RHO^1\, \MAT P_{10}) \otimes (\MAT P_{10}\, \RHO^2\, \MAT P_{10}) \big) \end{aligned} \\
=\M & \big( \begin{aligned}[t] & \!
\MAT Q \otimes \MAT Q \big) \odot \big( (\RHO^1 \otimes \RHO^2) (\MAT P_{11} \otimes \MAT P_{11}) + (\MAT P_{00} \otimes \MAT P_{10}) (\RHO^1 \otimes \RHO^2) (\MAT P_{11} \otimes \MAT P_{10}) ~\cdots \\
& \,+ (\MAT P_{10} \otimes \MAT P_{00}) (\RHO^1 \otimes \RHO^2) (\MAT P_{10} \otimes \MAT P_{11}) + (\MAT P_{10} \otimes \MAT P_{10}) (\RHO^1 \otimes \RHO^2) (\MAT P_{10} \otimes \MAT P_{10}) \big) \end{aligned} \\
\equiv\M & \MAT Q^{\otimes2} \odot \big( \RHO\, \MAT P_{33} + \MAT P_{10}\, \RHO\, \MAT P_{32} + \MAT P_{20}\, \RHO\, \MAT P_{31} + \MAT P_{30}\, \RHO\, \MAT P_{30} \big) ~, \label{eq:metamorph}
\end{split} \end{equation}
where $\RHO \equiv \RHO^1 \otimes \RHO^2$ and $\MAT Q^{\otimes N}$ ($N > 0$) denotes the $N$-fold Kronecker power of $\MAT Q$.
It is readily verified that the general formula for $N$ qubits is
\begin{equation} \label{eq:you}
\ALG U(\RHO) ~=~ \MAT Q^{\otimes N} \odot {\sum}_{m\,=\,0}^{\,M}\, \MAT P_{m,\M[0.05]0}\, \RHO\, \MAT P_{M,\M[0.05](M-m)} ~,
\end{equation}
where $\M[-0.05]\RHO\M[-0.05]$ is\M[-0.05] any \M[-0.05](not\M[-0.05] necessarily\M[-0.05] factorable)\M[-0.05] density\M[-0.05] matrix.\M[-0.05]
Similarly,\M[-0.05] the\M[-0.05] inverse\M[-0.05] is\M[-0.05] given\M[-0.05] by
\begin{equation} \label{eq:uoy}
\ALG U^{-1}(\SIG) ~=~ 2^{-N}\, {\sum}_{m\,=\,0}^{\,M}\, \MAT P_{m,\M[0.05]0}\, \big(\M[0.05] \OL{\MAT Q}^{\M[0.1]\otimes N\M[-0.1]} \odot \SIG \big)\, \MAT P_{M,\M[0.05](M-m)} ~.
\end{equation}
Evidently $\MAT Q^{\,\otimes N\!} \odot\. \ALG U^{-1}(\SIG) \,=\, 2^{-N}\. \ALG U(\. \OL{\MAT Q}^{\,\otimes N\!} \odot\. \SIG \.)$.

\section{Reality Check}
The following properties of the real density matrix $\SIG$ are worth noting explicitly:
\begin{itemize}
\item In addition to being real, it is nonsymmetric with one fixed element $\sigma_{00} = 1$;
\item It contains all the same information that the Hermitian density matrix does (since they are related by the bijection $\ALG U$).
\item It is diagonal if and only if the Hermitian density matrix is diagonal (which is why we defined $\sigma_{11}$ to be the coefficient of the diagonal Pauli matrix $\SIG[3]$).
\item It has the same tensor product structure as the Hermitian density matrix, since (as shown by Eq.~(\ref{eq:metamorph})) $\ALG U$ maps Kronecker products to Kronecker products.
\end{itemize}
In addition to these nice analytic features, the real density matrix can also be quite useful for displaying the results of \emph{quantum state tomography}: the determination of density matrices from experimental data.
In most cases to date, the real or imaginary parts of the Hermitian density matrix have been displayed using two-dimensional bar graphs (see e.g~\cite[\S7.7.4]{NielsChuan:00}).
Although useful, such a plot must both omit information and exhibit redundant information.
Real density matrices are definitely superior in this respect, and sometimes may also exhibit the underlying symmetry of a state more clearly.
Bar graphs of the real density matrix are shown below for both the diagonal Hilbert space basis as well as the Bell basis, illustrating how easily these states may be distinguished.
Further examples with experimental NMR data may be found in \citet{HCLBFPTWBH:02}.

\begin{figure}[h]
\begin{center} \begin{picture}(500,260)
\put(  0, 25){\includegraphics[width=100pt,height=100pt]{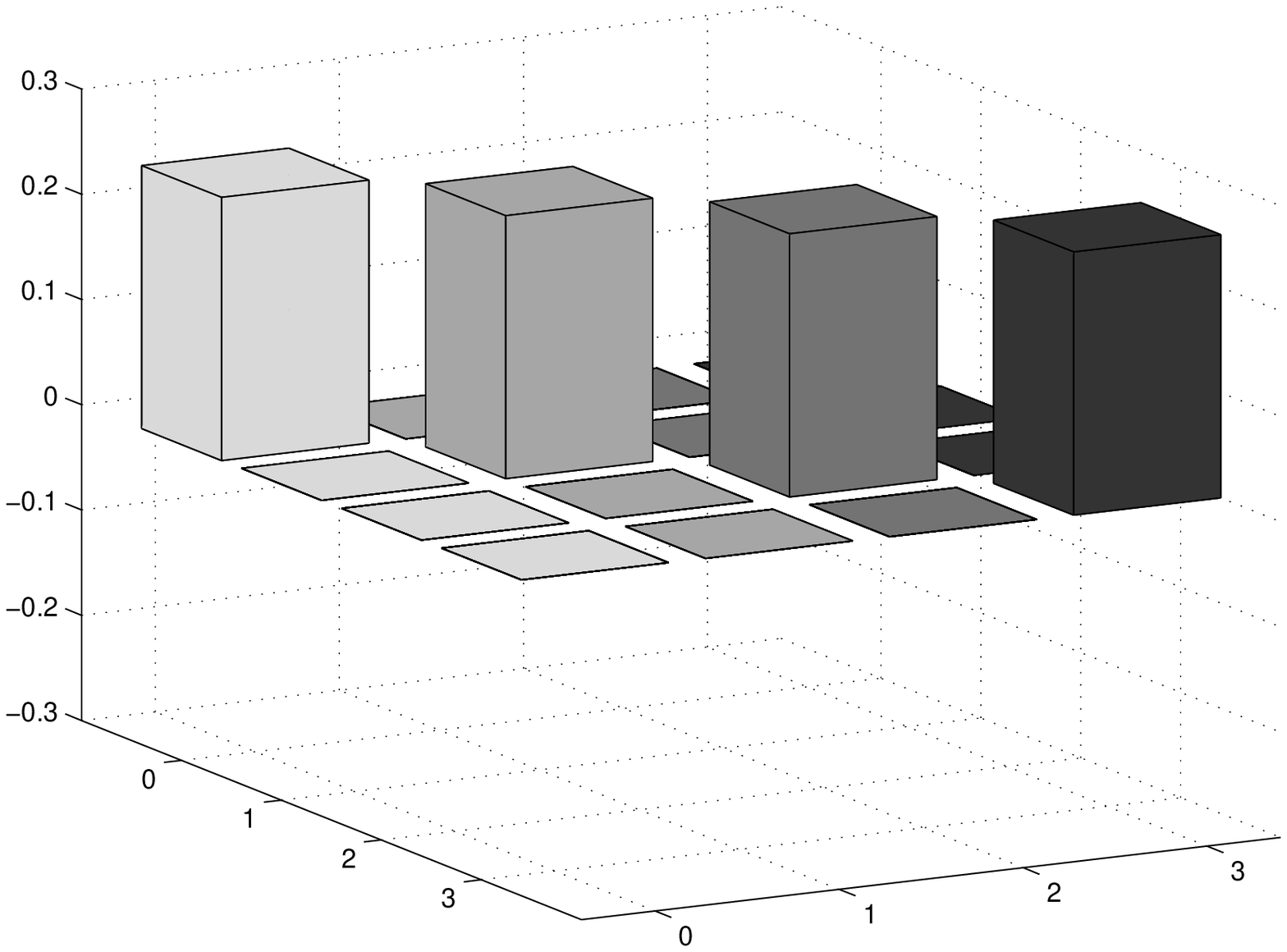}}
\put(125, 25){\includegraphics[width=100pt,height=100pt]{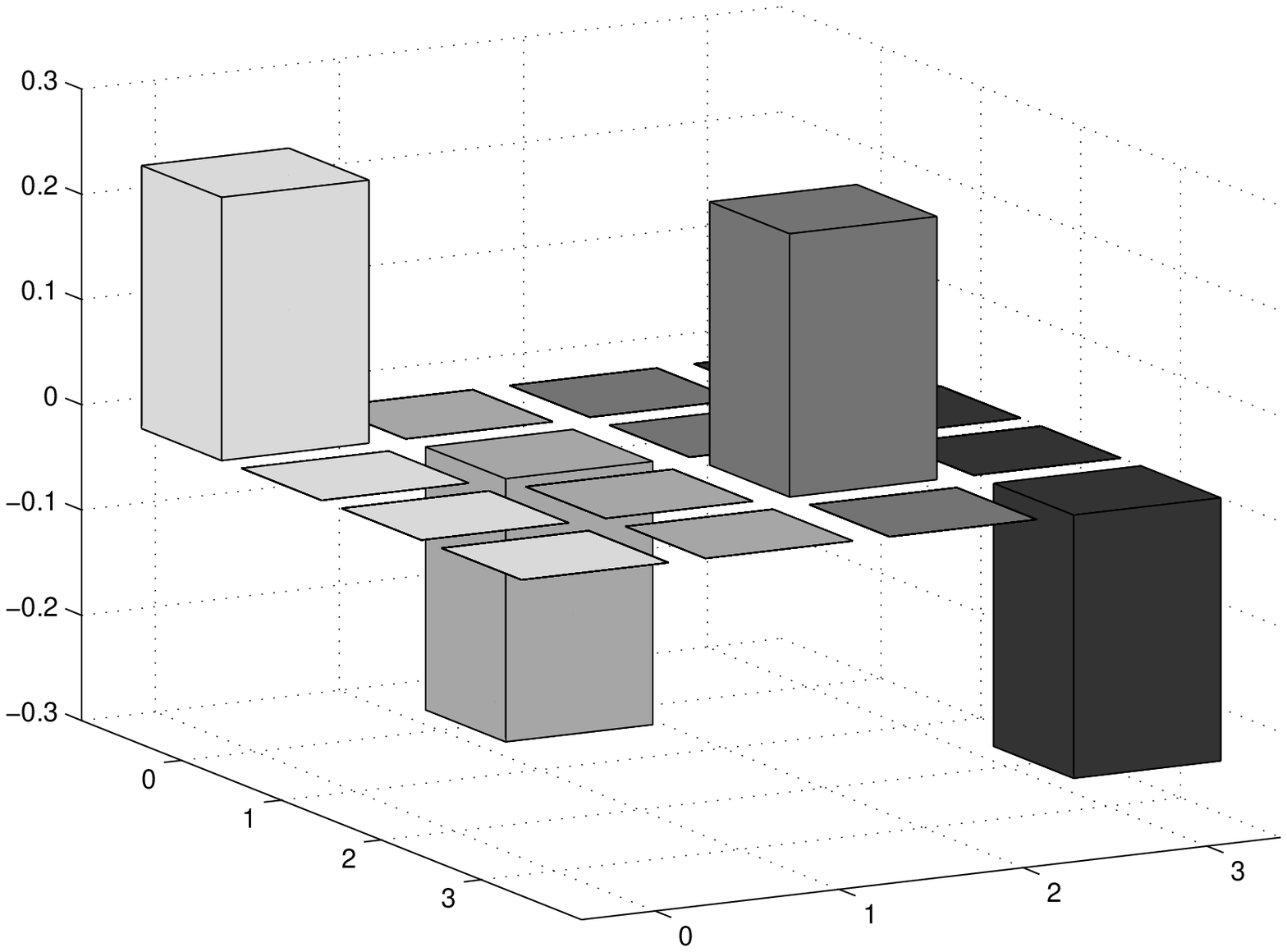}}
\put(250, 25){\includegraphics[width=100pt,height=100pt]{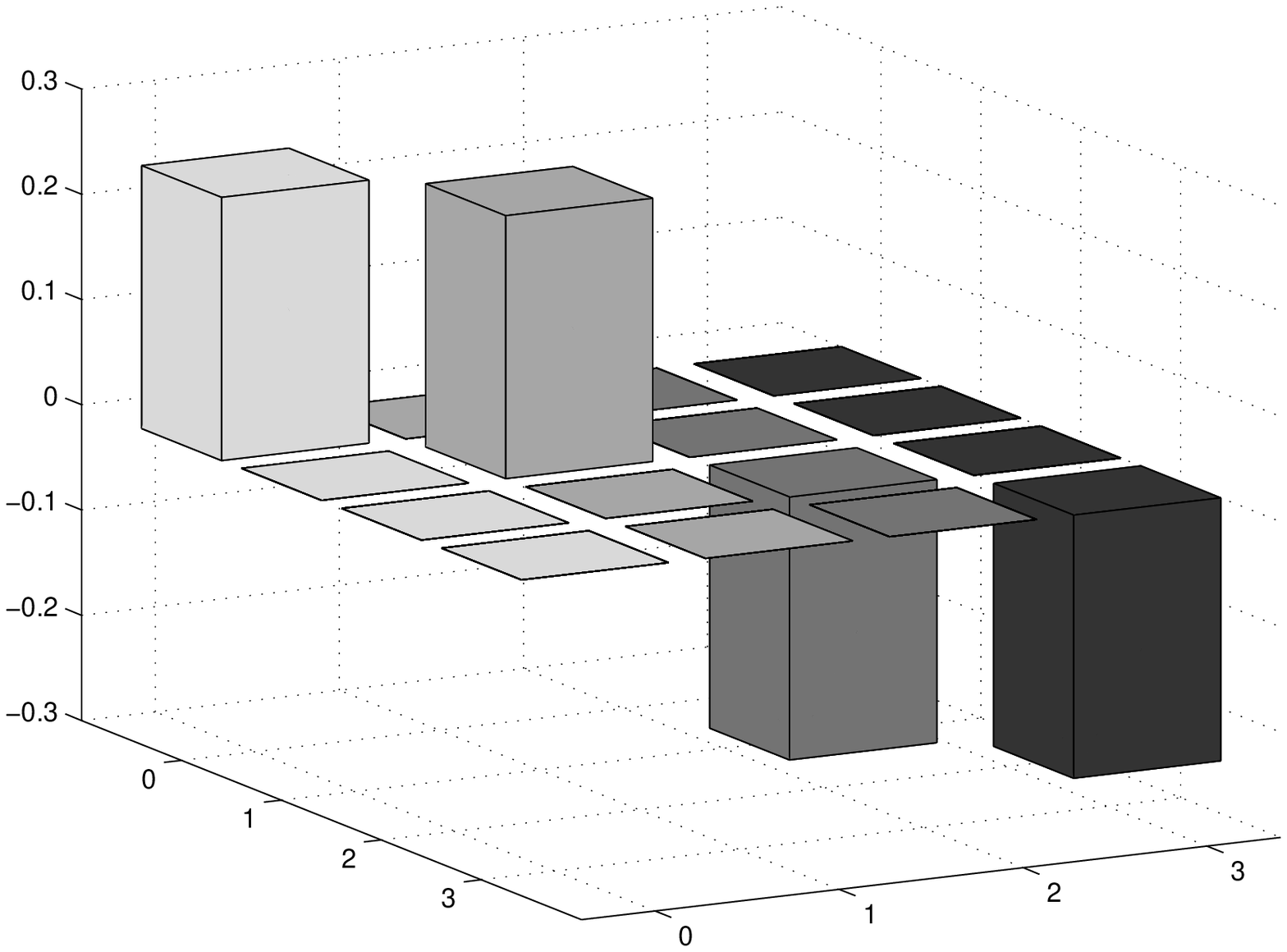}}
\put(375, 25){\includegraphics[width=100pt,height=100pt]{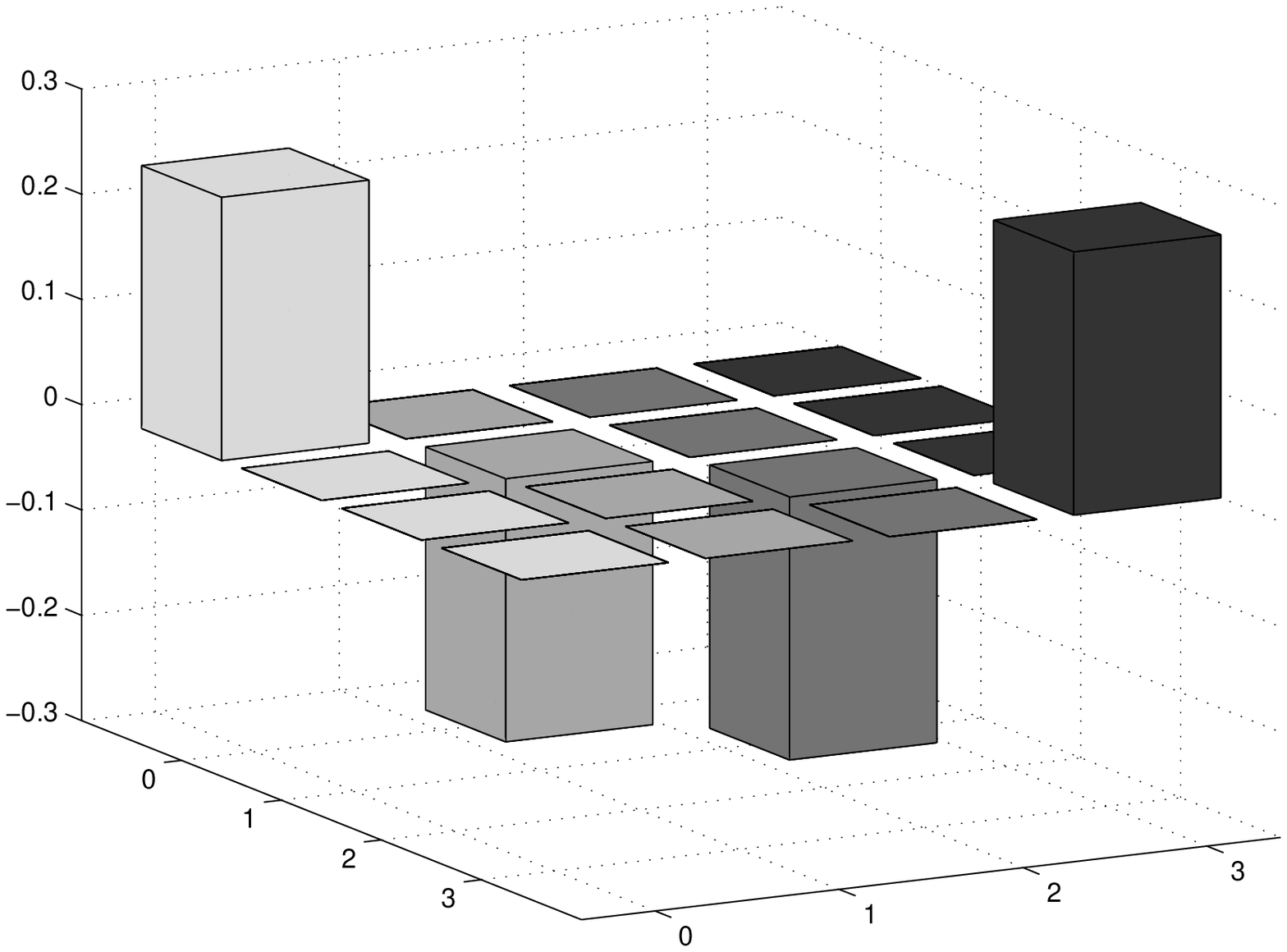}}
\put( 30, 10){$\KET{00}\BRA{00}$}
\put(155, 10){$\KET{01}\BRA{01}$}
\put(280, 10){$\KET{10}\BRA{10}$}
\put(405, 10){$\KET{11}\BRA{11}$}
\put(  0,155){\includegraphics[width=100pt,height=100pt]{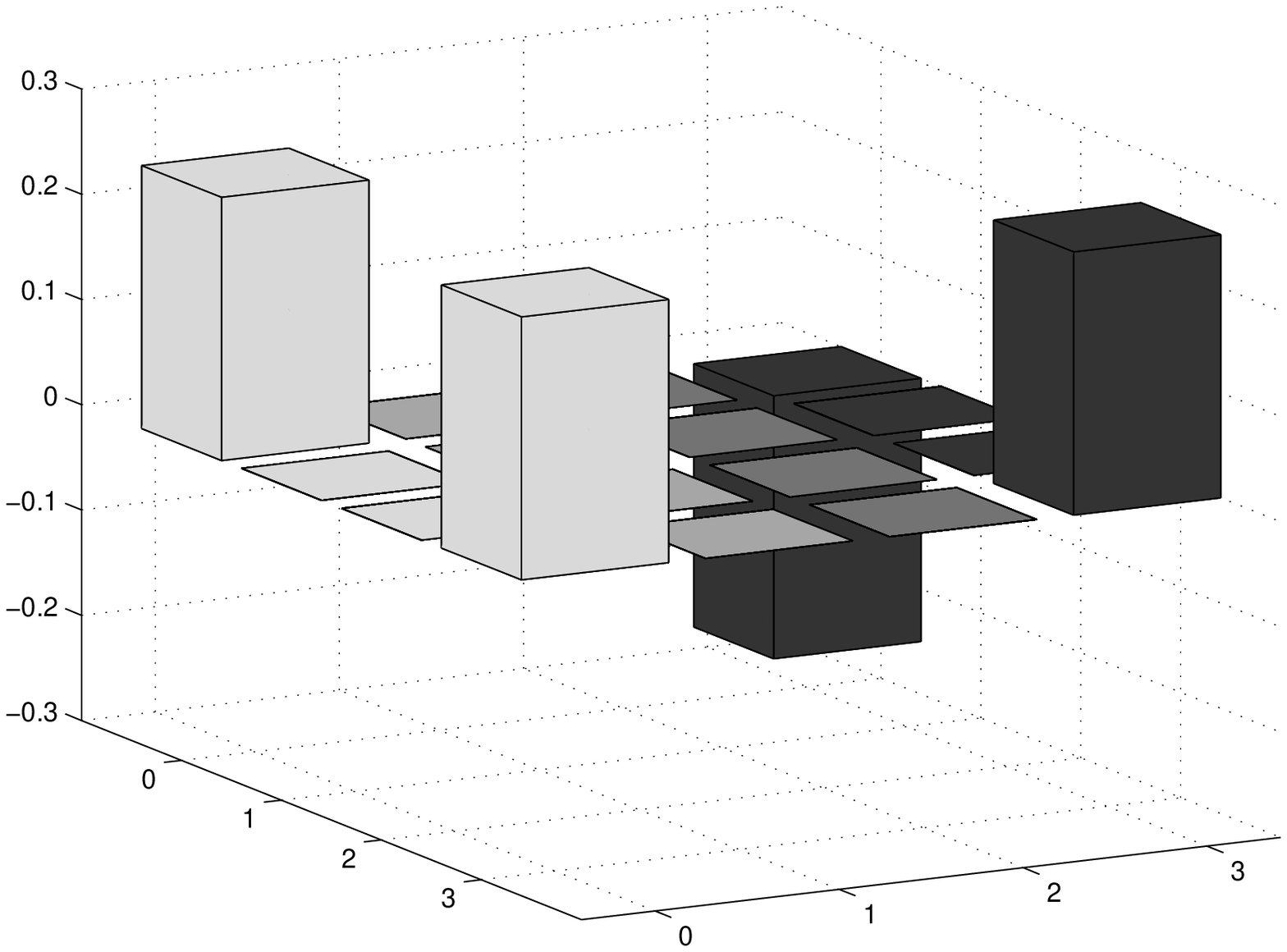}}
\put(125,155){\includegraphics[width=100pt,height=100pt]{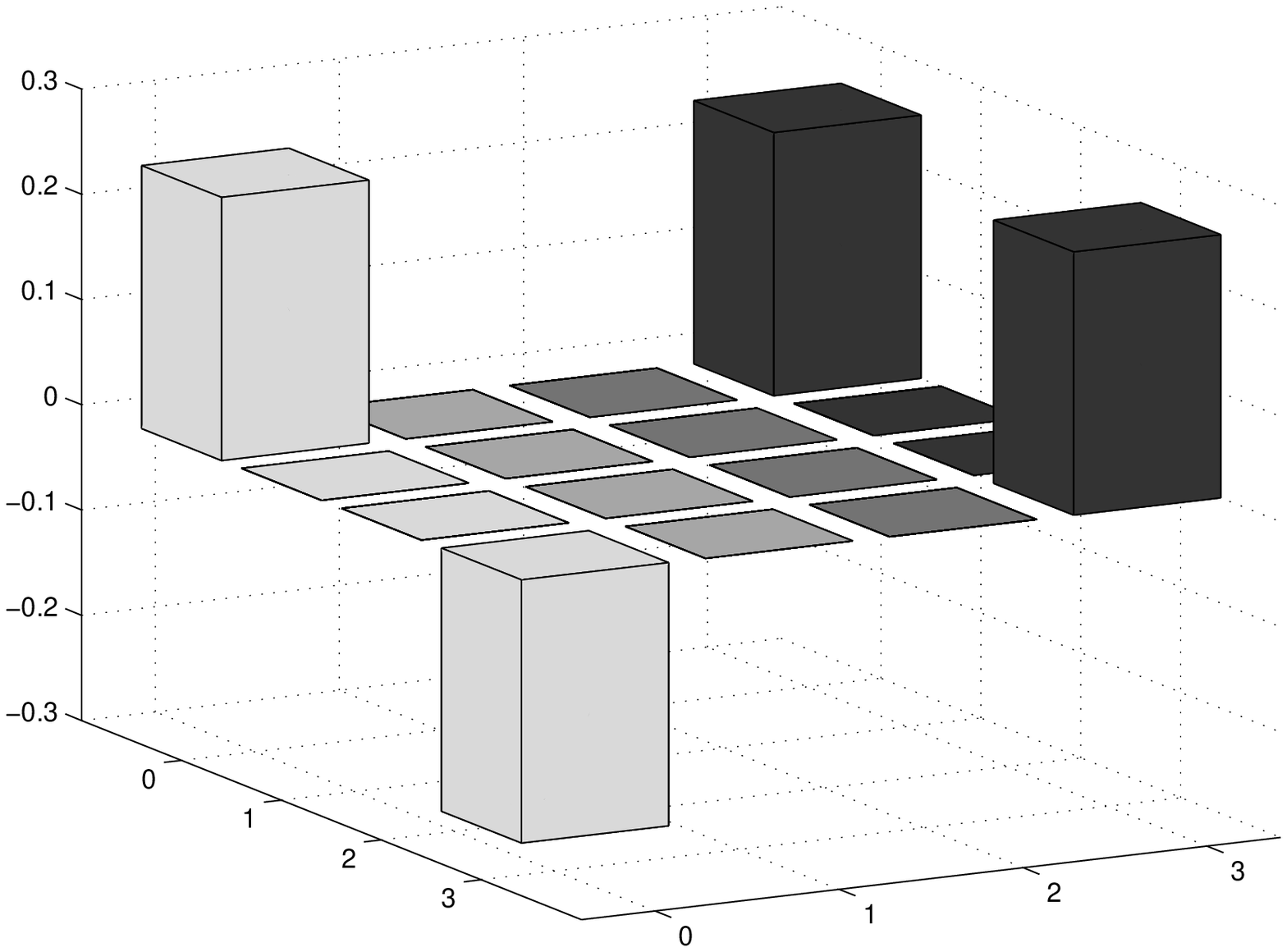}}
\put(250,155){\includegraphics[width=100pt,height=100pt]{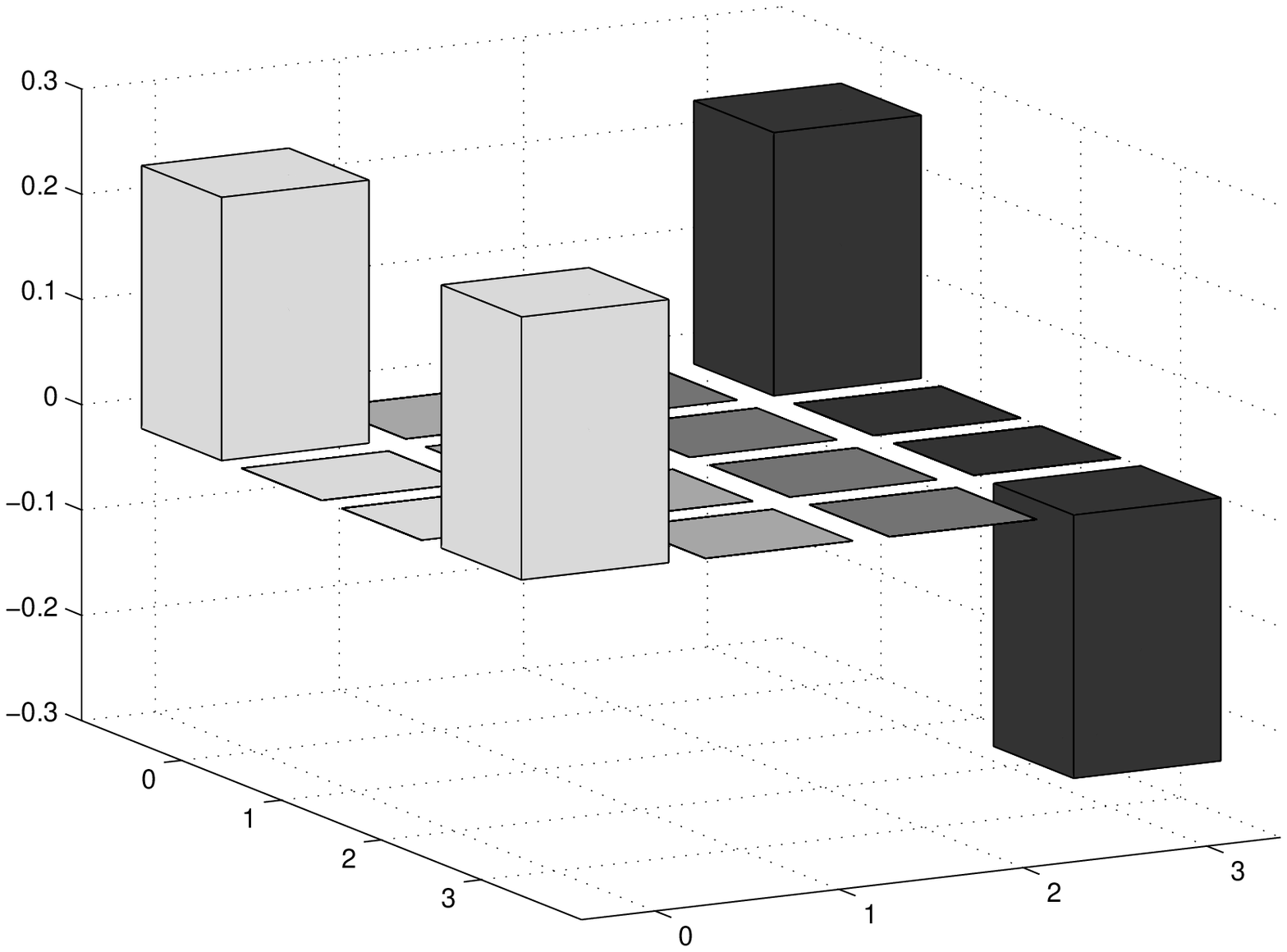}}
\put(375,155){\includegraphics[width=100pt,height=100pt]{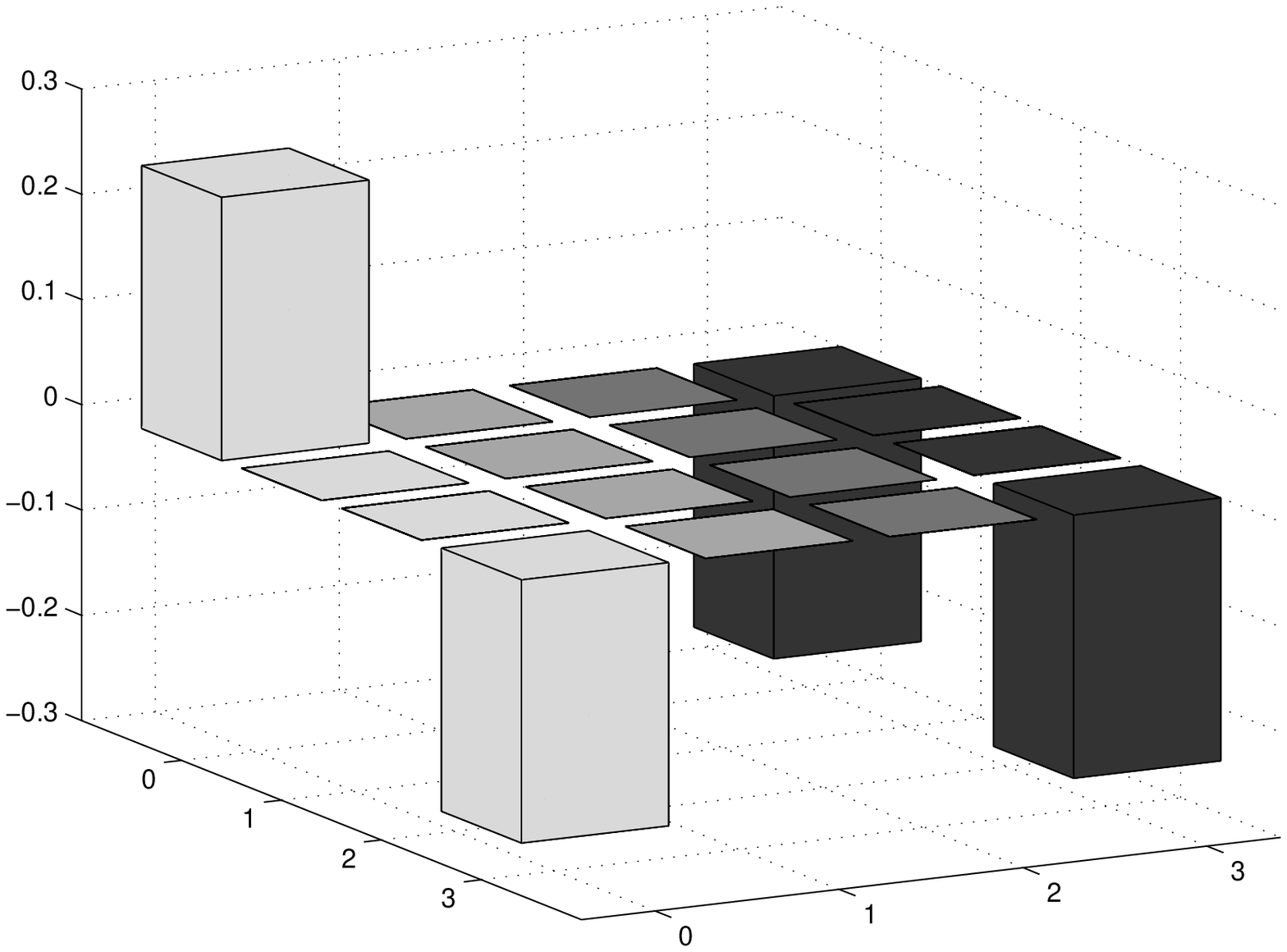}}
\put(  5,140){$\KET{00}\BRA{00} + \KET{11}\BRA{11}$}
\put(130,140){$\KET{00}\BRA{00} - \KET{11}\BRA{11}$}
\put(255,140){$\KET{01}\BRA{01} + \KET{10}\BRA{10}$}
\put(380,140){$\KET{01}\BRA{01} - \KET{10}\BRA{10}$}
\end{picture}
\caption{Plots of the diagonal Hilbert space basis (below) and of the Bell basis (above).
All axes are dimensionless, and the labels on the horizontal axes correspond to the indices of the two-qubit real density matrix entries $\sigma_{ij}$ as used in the main text.}
\end{center}
\end{figure}

The fact that the mapping between Pauli matrix coefficients and the entries of the Hermitian density operator preserves the tensor product structure has been noted earlier by \citet{PitteRubin:00a} (and without doubt by many other researchers as well).
In our present notation, their observation was based upon the following simple relation:
\begin{equation}
\begin{bmatrix} \sigma_{00} & \sigma_{10} \\ \sigma_{11} & \imath\sigma_{01} \end{bmatrix} ~=~ \begin{bmatrix} ~1~ & ~1~ \\ ~1~ & -1~ \end{bmatrix} \begin{bmatrix} \rho_{00} & \rho_{10} \\ \rho_{11} & \rho_{01} \end{bmatrix} ~.
\end{equation}
While this is certainly a simpler relation than our operator sum, the ``density matrix'' on the right-hand side is not the usual Hermitian one, and the mapping between the two can be written explicitly only by using operator sums, supermatrices, or the like.
Our goal here is to translate the usual operations and relations on Hermitian density matrices into the real domain, and the reordering of the entries of the real density matrix as above offers no advantage for this purpose.

As our first example of such a translation, let us show how the usual criterion for the purity of the Hermitian density matrix can be carried over to the real domain:
\begin{align}
\M[-3] 1 \M=\M 2^N\, \langle\, \RHO^2\, \rangle \M=\M &
\left\langle\, \RHO \,{\sum}_{m\,=\,0}^{m=M}\, \MAT P_{m,0} \big( \OL{\MAT Q}^{\,\otimes N} \odot \SIG \big) \MAT P_{M,M-m}\right\rangle \notag  \\ =\M &
\left\langle\! \big( \OL{\MAT Q}^{\,\otimes N} \odot \SIG \big) {\sum}_{m=0}^M\, \MAT P_{M,M-m}\, \RHO\, \MAT P_{m,0} \right\rangle \notag \\ =\M &
\left\langle\, {\sum}_{m\,=\,0}^{m=M}\, \MAT P_{m,0}\, \RHO\, \MAT P_{M,M-m}\, \big( \OL{\MAT Q}^{\,\otimes N} \odot \SIG \big)^{\!\dag} \right\rangle \\ \notag =\M &
\left\langle\! \big( \OL{\MAT Q}^{\,\otimes N} \odot \SIG \big) \big( \OL{\MAT Q}^{\,\otimes N} \odot \SIG \big)^{\!\dag} \right\rangle \\ \notag =\M &
\left\langle \big( \MAT Q^{\otimes N} \odot \OL{\MAT Q}^{\,\otimes N} \odot \SIG \big)^{\dag}\, \SIG\, \right\rangle ~=~ \big\langle\, \SIG^\top \SIG\, \big\rangle ~.
\end{align}
In going to the last line, we have used the general relation $\langle (\MAT A \odot \MAT B)\, \MAT C^\dag \rangle = \langle (\MAT A \odot \MAT C)^\dag\, \MAT B\. \rangle$ for arbitrary conformant matrices $\MAT A, \MAT B, \MAT C$ \cite{Lutkepohl:96}.
This derivation easily generalizes to a formula for the ensemble-average expectation values of any observable with Hermitian matrix $\EMB\mu$ and corresponding real matrix $\EMB\nu = \ALG U(\EMB\mu)$, showing that
\begin{equation}
\big\langle\, \EMB\mu \,\big|\, \RHO\, \big\rangle ~\equiv~ 2^{\.N}\, \big\langle\, \EMB\mu\, \RHO\, \big\rangle ~=~\big\langle\, \EMB\nu^\top \SIG\, \big\rangle ~=~ 2^{\.-N\.} \big\langle\, \EMB\nu \,\big|\, \SIG \,\big\rangle ~.
\end{equation}

We close this section by noting that the \emph{partial trace} operation corresponds simply to extracting a principal submatrix of the real density matrix \cite{SomCorHav:98}.

\section{Life in the Real World}
While the expectation values of observables carry over to the real domain without significant complication, things become distinctly more challenging when it comes to integrating the equations of motion.
In the case of a single qubit, it is readily verified that the commutator with an arbitrary Hamiltonian $\EMB\mu = \ALG U^{-1}(\EMB\nu)$ becomes
\begin{equation} \label{eq:1pc} \begin{split}
\big[\M[-0.25]\big[\, \SIG,\, \EMB\nu\, \big]\M[-0.25]\big] ~\equiv~ \ALG U\big( \big[\, \RHO,\, \EMB\mu\, \big] \big)
& ~=\M[.5] \imath\,\MAT P_{01} \big( \EMB\nu\, \MAT E_{11}\M[.1] \SIG \,-\,\SIG\, \MAT E_{11}\M[.1] \EMB\nu \big) \MAT P_{01} \\
& ~=\M[.5] \imath \begin{bmatrix} 0 & \sigma_{11\,} \nu_{10} -\sigma_{10\,} \nu_{11} \\ \sigma_{01\,} \nu_{11} - \sigma_{11\,} \nu_{01} & \sigma_{10\,} \nu_{01} - \sigma_{01\,} \nu_{10} \end{bmatrix} ~,
\end{split} \end{equation}
wherein the matrix entries are the components of the usual vector cross product.

This equation of motion is most simply integrated by considering the matrix representation of the commutation superoperator defined by $\EMB\nu$, which we henceforth assume without loss of generality has $\nu_{00} = 0$.
Letting $\KET{\MAT X}$ denote the column vector of height $(M+1)^2$ obtained by stacking the columns of the $(M+1)\times(M+1)$ matrix $\MAT X$ on top of one another in left-to-right order, and applying the well-known identity
\begin{equation} \label{eq:oppr2prop}
\KET{ \MAT{AXB} } ~=~ \big( \MAT B^\top \otimes \MAT A \big)\, \KET{\MAT X}
\end{equation}
(see e.g.~\cite{Lutkepohl:96}), we find that\footnote{%
The factor of $1/2$ here does not mean that the rotation is spinorial, but rather that the rate of rotation is $1/\sqrt2$ times the Hilbert-Schmidt norm of the Pauli matrix that generates it, while we pick up another factor of $1/\sqrt2$ on transforming to the real domain.}
\begin{equation} \label{eq:form0} \begin{split}
\M[-1] \dot{\SIG} ~=~ \Big|\, \tfrac{\displaystyle\imath}2\, \big[\M[-0.25]\big[\, \SIG,\, \EMB\nu\, \big]\M[-0.25]\big] \Big\rangle ~=\M[0.5]
& \text{\large$\HALF$}\, \Big( \MAT P_{01} \otimes \MAT P_{01} \Big) \Big( \MAT P_{00} \otimes \EMB\nu\, \MAT E_{11} \,-\,\EMB\nu^\top \MAT E_{11} \otimes \MAT P_{00} \Big)\, \big|\, \SIG\, \big\rangle \\ =\M[0.5] 
& \frac12 \begin{bmatrix} ~0&0&0&0\\ ~0&0&-\nu_{11}&\nu_{01}\\ ~0&\nu_{11}&0&-\nu_{10}\\ ~0&-\nu_{01}&\nu_{10}&0 \end{bmatrix}\! \begin{bmatrix} \sigma_{00}\\ \sigma_{10}\\ \sigma_{01}\\ \sigma_{11} \end{bmatrix} ~\equiv~ \HALF\, \EMB{\ALG R}_{\EMB\nu}\, \KET{\SIG}\, .
\end{split} \end{equation}
Since $\EMB{\ALG R}_{\EMB\nu}^{\,3} = -{\|\EMB\nu\|}^{2\,} \EMB{\ALG R}_{\EMB\nu\,}$ ($\|\EMB\nu\|^2 \equiv 2\M[0.05] \langle \EMB\nu^\top \EMB\nu \rangle \equiv \HIP{\EMB\nu}{\EMB\nu}$), this one-sided matrix differential equation is easily integrated by using the Cayley-Hamilton theorem to exponentiate the (lower-right $3\times3$ block of the) coefficient matrix \cite{NajfeHavel:95a}, obtaining
\begin{equation} \begin{split}
& \big|\M[0.1]\SIG(t)\M[0.1]\big\rangle \M[.5]=\M[.5] \bigg( \MAT P_{00} \otimes \MAT P_{00} \,+\, \frac{\sin(\|\EMB\nu\|\M[.1]t/2\M[.1])}{\|\EMB\nu\|}\, \EMB{\ALG R}_{\EMB\nu} \,+\, \frac{1-\cos(\|\EMB\nu\|\M[.1]t/2\M[.1])} {\|\EMB\nu\|{\X[1.4]}^2}\, \EMB{\ALG R}_{\EMB\nu}^{\,2} \bigg)\, \big|\M[0.1]\SIG(0)\M[0.1]\big\rangle \\ =\M[.5]
& \Big|\, \SIG(0) \,+\, \imath\, \sin(\|\EMB\nu\|\M[.1]t/2\M[.1])\, \big[\M[-0.25]\big[\, \SIG(0),\, \hat{\EMB\nu}\, \big]\M[-0.25]\big] \,-\, \big( 1-\cos(\|\EMB\nu\|\M[.1]t/2\M[.1]) \big)\, \big[\M[-0.25]\big[ \big[\M[-0.25]\big[\, \SIG(0),\, \hat{\EMB\nu}\, \big]\M[-0.25]\big], \hat{\EMB\nu}\, \big]\M[-0.25]\big] \Big\rangle ,
\end{split} \end{equation}
wherein $\hat{\EMB\nu} \equiv \EMB\nu/\|\EMB\nu\|$ and it is readily shown that
\begin{equation} \label{eq:r2}
\EMB{\ALG R}_{\EMB\nu}^{\,2\,} ~=~ \KET{\EMB\nu} \BRA{\EMB\nu} ~-~ \|\EMB\nu\|^2\M[0.25] \big( \MAT P_{00} \otimes \MAT P_{00} \:-\: \MAT E_{00} \otimes \MAT E_{00} \big)
\end{equation}
is $-\|\EMB\nu\|^2$ times the projection onto the plane orthogonal to the unit vector $\KET{\hat{\EMB\nu}} =  \BRA{\hat{\EMB\nu}}^\top$.
The whole formula can thus be expressed more geometrically as
\begin{equation}
\KET{\VP\SIG(t)\!} ~=~ \HIP{\hat{\EMB\nu}}{\VP\SIG}\; \KET{\hat{\EMB\nu}} \,+\, \sin(\|\EMB\nu\|\M[.1]t/2\M[.1])\, \KET{\hat{\EMB\nu}} \times \KET{\VP{\SIG}} \,+\, \cos(\|\EMB\nu\|\M[.1]t/2\M[.1])\, \KET{\hat{\EMB\nu}} \times \big(\. \KET{\hat{\EMB\nu}} \times \KET{\VP{\SIG}} \big) ,~
\end{equation}
where $\VP{\SIG} \equiv \SIG(0) - \MAT E_{00}$ and ``$\times$'' is the cross product of the (last three components of the) vectors it connects.
This is a standard expression for rotation of the three-dimensional vector $\KET{\VP\SIG}$ about the axis $\KET{\hat{\EMB\nu}}$ by an angle $\|\EMB\nu\|\M[.1]t/2\M[0.05]$.

The extension of these formulae to general multi-particle commutators is not straightforward, since the tensor product of commutation superoperators is not simply related to the tensor products of their underlying operators.
Nevertheless, we can give a reasonably simple formula for the two-particle commutator with a factorable Hamiltonian, which is the most important case in practice.
This is based upon a geometric algebra expression for the commutator of a tensor product of two three-dimensional vectors (or equivalently in the present context, traceless $2\times2$ Hermitian matrices), which is derived in \citet{HaDo$AGACSE:02}:
\begin{equation} \label{eq:doran}
2\, \big[ \MAT A \otimes \MAT C,\, \MAT B \otimes \MAT D \big] ~=~ \HIP{\MAT A}{\MAT B} \big( \MAT P_{00} \otimes [ \MAT C, \MAT D ] \big) ~+~ \big( [ \MAT A, \MAT B ] \otimes \MAT P_{00} \big) \HIP{\MAT C}{\MAT D} ~.
\end{equation}
Letting $\MAT a \equiv \ALG U(\MAT A)$, etc.~be the corresponding real matrices, this translates to:
\begin{equation} \label{eq:form1a}
2\, \big[\M[-0.25]\big[ \MAT a \otimes \MAT c ,\, \MAT b \otimes \MAT d \big]\M[-0.25]\big] ~=~ \HIP{\MAT a}{\MAT b} \big( \MAT E_{00} \otimes \M[.1][\M[-.15][\. \MAT c, \MAT d \M[.1]]\M[-.15]] \big) ~+~ \big( \M[.1][\M[-.15][\. \MAT a, \MAT b \M[.1]]\M[-.15]] \otimes \MAT E_{00} \big)\HIP{\MAT c}{\MAT d}
\end{equation}
This formula is easily extended to the case in which $a_{00},\ldots,d_{00} \ne 0$ by multilinearity; in the following, however, we will need only the case in which $a_{00} = c_{00} = 1$, which introduces two additional terms:
\begin{equation} \label{eq:form1b} \begin{split}
\M[-.5] \big[\M[-0.25]\big[\. \MAT E_{00} \otimes \MAT c ,\, \MAT b \otimes \MAT d \.\big]\M[-0.25]\big] \M[.5]\leftrightarrow\M[.5] \HALF\, \big[\. \MAT P_{00} \otimes \MAT C \,,\: \MAT B \otimes \MAT D \big] ~=~ \HALF\, \MAT B \otimes [\M[.1] \MAT C , \MAT D \M[.1]] \M[.5]\leftrightarrow\M[.5]
& \HALF\, \MAT b \otimes [\M[-.15][\M[.1] \MAT c, \MAT d \M[.1]]\M[-.15]] ~;
\\
\M[-.5] \big[\M[-0.25]\big[\. \MAT a \otimes \MAT E_{00} ,\, \MAT b \otimes \MAT d \.\big]\M[-0.25]\big] \M[.5]\leftrightarrow\M[.5] 
\HALF\, \big[ \MAT A \otimes \MAT P_{00} \,,\: \MAT B \otimes \MAT D \big] ~=~ \HALF\, [\M[.1] \MAT A ,\, \MAT B \M[.1]] \otimes \MAT D \M[.25]\leftrightarrow\M[.5]
& \HALF\, [\M[-.15][\M[.1] \MAT a, \MAT b \M[.1]]\M[-.15]] \otimes \MAT d ~.
\end{split} \end{equation}
Finally, we shall need the general result \cite{Lutkepohl:96,Havel!QPT:03}:
\begin{equation} \label{eq:form2}
\KET{ \MAT X \otimes \MAT Y } ~=~ \big( \MAT P_{00} \otimes \EMB{\ALG K}_{22} \otimes \MAT P_{00} \big)\, \big(\, \KET{ \MAT X } \otimes \KET{ \MAT Y } \big) ,
\end{equation}
for $2\times2$ matrices $\MAT X$ \& $\MAT Y$, where the two-particle \emph{commutation matrix} $\EMB{\ALG K}_{22}$ is given by
\begin{equation}
\EMB{\ALG K}_{22} ~=~ {\sum}_{i,\,j=0}^{\,3,\,3}\, \MAT E_{ij} \otimes \MAT E_{ji} ~=~ \begin{bmatrix} ~1~&~0~&~0~&~0~\\[-1ex] ~0~&~0~&~1~&~0~\\[-1ex] ~0~&~1~&~0~&~0~\\[-1ex] ~0~&~0~&~0~&~1~ \end{bmatrix} ~.
\end{equation}

We are interested in the case that $\MAT a = \SIG^1$, $\MAT b = \EMB\nu^1$, $\MAT c = \SIG^2$ and $\MAT d = \EMB\nu^2$, i.e.~we have a factorizable two-particle state $\SIG^1 \otimes \SIG^2$ evolving under a bi-axial interaction $\EMB\nu^1 \otimes \EMB\nu^2$.
To express this more compactly, we define the matrix $\EMB{\ALG S}_{\EMB\nu^1}$ via
\begin{equation}
\big\langle\, \EMB\nu^1\, \big|\, \SIG^1\, \big\rangle\, \big|\, \MAT E_{00}\, \big\rangle ~+~ \big|\, \EMB\nu^1\, \big\rangle ~=~ \begin{bmatrix} ~0~&\nu_{10}^1&\nu_{01}^1&\nu_{11}^1\\[-1ex] \nu_{10}^1&0&0&0\\[-1ex] \nu_{01}^1&0&0&0\\[-1ex] \nu_{11}^1&0&0&0 \end{bmatrix}\! \begin{bmatrix} \text{\small$1$}\\[-1ex] \sigma_{10}^1\\[-1ex] \sigma_{01}^1\\[-1ex] \sigma_{11}^1 \end{bmatrix} ~\equiv\M \EMB{\ALG S}_{\!\EMB\nu^1}\, \big|\, \SIG^1\, \big\rangle
\end{equation}
with an analogous definition in $\HIP{\EMB\nu^2}{\SIG^2}\, \KET{\MAT E_{00}} + \KET{\EMB\nu^2} \equiv\, \EMB{\ALG S}_{\!\EMB\nu^2}\, \KET{\SIG^2}$.
Then Eqs.~(\ref{eq:form0}), (\ref{eq:form1a}), (\ref{eq:form1b}) \& (\ref{eq:form2}) give us
\begin{align} \label{eq:ugly}
\M[-1]\partial_t\, \big|\, \SIG^1\otimes\SIG^2\, \big\rangle \M[.25]=\M[.5]
&\notag \Big|\M[0.2] \tfrac{\displaystyle\imath}4\M[0.1] \big[\M[-.25]\big[ \SIG^1 \otimes \SIG^2 ,\: \EMB\nu^1 \otimes \EMB\nu^2\, \big]\M[-.25]\big] \Big\rangle \\ =\M[.5]
& \big( \MAT P_{00} \otimes \EMB{\ALG K}_{22} \otimes \MAT P_{00} \big)\M[0.1]  \tfrac{\displaystyle\imath}4
\begin{aligned}[t] \Big(
& \HIP{\SIG^1}{\EMB\nu^1}\, \big|\, \MAT E_{00}\, \big\rangle \otimes \big|\M[0.1] \big[\M[-.25]\big[ \SIG^2,\: \EMB\nu^2\, \big]\M[-.25] \big] \big\rangle ~+ \\
\M[-3] \cdots~ & \big|\M[0.1] \big[\M[-.25]\big[ \SIG^1,\: \EMB\nu^1\, \big]\M[-.25]\big] \big\rangle \otimes \big|\, \MAT E_{00}\, \big\rangle\, \HIP{\SIG^2}{\EMB\nu^2} ~+ \\
\M[-3] \cdots~ & \big|\, \EMB\nu^1\, \big\rangle \otimes \big|\M[0.1] \big[\M[-.25]\big[ \SIG^2 ,\: \EMB\nu^2 \big]\M[-.25]\big] \big\rangle ~+ \\
\M[-3] \cdots~ & \big|\M[0.1] \big[\M[-.25]\big[ \SIG^1 ,\: \EMB\nu^1 \big]\M[-.25]\big]\, \big\rangle \otimes \big|\, \EMB\nu^2\, \big\rangle \Big)
\end{aligned}
\\ =\M[.5]
&\notag \big( \MAT P_{00} \otimes \EMB{\ALG K}_{22} \otimes \MAT P_{00} \big) \,\tfrac14\, \big( \EMB{\ALG S}_{\EMB\nu^1} \otimes \EMB{\ALG R}_{\EMB\nu^2}
\begin{aligned}[t]
\,+\, \EMB{\ALG R}_{\EMB\nu^1} \otimes \EMB{\ALG S}_{\EMB\nu^2} \big)
\big(\M[0.1] \big|\M[0.1] \SIG^1\M[0.1] \big\rangle \otimes \big|\M[0.1] \SIG^2\M[0.1] \big\rangle \big) 
\end{aligned}
\\ =\M[.5]
&\notag \big( \MAT P_{00} \otimes \EMB{\ALG K}_{22} \otimes \MAT P_{00} \big) \,\tfrac14\,
\begin{aligned}[t]
\big( \EMB{\ALG S}_{\EMB\nu^1} \otimes \EMB{\ALG R}_{\EMB\nu^2} \,+\, \EMB{\ALG R}_{\EMB\nu^1} \otimes \EMB{\ALG S}_{\EMB\nu^2}  \big) ~\cdots \M[4] & \\
\cdots~ \big( \MAT P_{00} \otimes \EMB{\ALG K}_{22} \otimes \MAT P_{00} \big)\, \big|\,  \SIG^1 \otimes \SIG^2\, \big\rangle & ~.
\end{aligned}
\end{align}

Since left or right multiplication of a $4$D column or row vector by $\EMB{\ALG R}_{\MAT x}$ gives the cross product of the last three components of that vector with $[ x_{10}, x_{01}, x_{11} ]$, it may be seen that $\EMB{\ALG S}_{\EMB\nu^1}$, $\EMB{\ALG R}_{\EMB\nu^1}$ are mutually annihilating (i.e.~$\EMB{\ALG S}_{\EMB\nu^{1\,}} \EMB{\ALG R}_{\EMB\nu^1} = \EMB{\ALG R}_{\EMB\nu^1\,} \EMB{\ALG S}_{\EMB\nu^1} = \MAT 0$), and similarly for $\EMB{\ALG S}_{\EMB\nu^2}$, $\EMB{\ALG R}_{\EMB\nu^2}$.
As a result, the two terms on the last line of Eq.~(\ref{eq:ugly}) commute and their exponential factorizes.
It is moreover easily shown that $\EMB{\ALG S}_{\MAT x}^{\M[.1]3} = \|\MAT x\|^2 \EMB{\ALG S}_{\X[1.5]\MAT x}$ and $\EMB{\ALG R}_{\MAT x}^3 = -\|\MAT x\|^{2\,} \EMB{\ALG R}_{\X[1.5]\MAT x\,}$, so the overall integral is
\begin{equation} \label{eq:overall} \begin{split}
\big|\M[0.1] \SIG(t)\M[0.1] \big\rangle ~=~ \big( \MAT P_{00} \otimes \EMB{\ALG K}_{22} \otimes \MAT P_{00} \big)\, \MAT{Exp}\big( \EMB{\ALG S}_{\EMB\nu^1} \otimes \EMB{\ALG R}_{\EMB\nu^2} \M[0.25]t/4 \big)\, \MAT{Exp}\big( \EMB{\ALG R}_{\EMB\nu^1} \otimes \EMB{\ALG S}_{\EMB\nu^2} \M[0.25]t/4 \big) ~\cdots \M[2] & \\
\cdots~ \big( \MAT P_{00} \otimes \EMB{\ALG K}_{22} \otimes \MAT P_{00} \big)\, \big|\M[0.1]  \SIG(0)\M[0.1] \big\rangle , &
\end{split} \end{equation}
where $\SIG(0) = \SIG^1 \otimes \SIG^2$ and (letting $\MAT P_{00}^{\otimes4}$ be the $2\times2$ identity tensored with itself $4$ times)
\begin{equation} \begin{split}
\M[-1] \MAT{Exp}\big(\,\EMB{\ALG S}_{\EMB\nu^1} \otimes \EMB{\ALG R}_{\EMB\nu^2} \M[.25]t/4 \big) \M[.5]=\M[.5]
\MAT P_{00}^{\otimes4} \,+\, \frac{\sin\!\big( \|\EMB\nu^1\| \|\EMB\nu^2\| \M[0.1]t/4 \big)}{\|\EMB\nu^1\| \|\EMB\nu^2\|}\, \big( \EMB{\ALG S}_{\EMB\nu^1} \otimes \EMB{\ALG R}_{\EMB\nu^2} \big) ~+ &
\\ \cdots~ \frac{1 - \cos\!\big(\|\EMB\nu^1\| \|\EMB\nu^2\| \M[0.1]t/4 \big)}{{\|\EMB\nu^1\|}^2 {\|\EMB\nu^2\|}^2}\, \big( \EMB{\ALG S}_{\EMB\nu^1} \otimes \EMB{\ALG R}_{\EMB\nu^2} \big)^2 &
\end{split} \end{equation}
with an almost identical expression for $\MAT{Exp}\big( \EMB{\ALG R}_{\EMB\nu^1} \otimes \EMB{\ALG S}_{\EMB\nu^2} \M[.25]t/4 \big)$.
The square in the last term may be evaluated by combining Eq.~(\ref{eq:r2}) with
\begin{equation} \label{eq:s2}
\EMB{\ALG S}_{\MAT x}^{\M[.1]2} ~=~ \KET{\MAT x} \BRA{\MAT x} ~+~ \|\MAT x\|^2\M[0.25] \MAT E_{00} \otimes \MAT E_{00} ~.
\end{equation}

Because $\EMB{\ALG S}_{\EMB\nu^1} \otimes \EMB{\ALG R}_{\EMB\nu^2}$ and $\EMB{\ALG R}_{\EMB\nu^1} \otimes \EMB{\ALG S}_{\EMB\nu^2}$ are mutually annihilating, the product of their exponentials expands to only five terms, two pairs of which have identical trigonometric coefficients.
On pulling the right-hand column operator ``$\KET{}$'' back to the left in Eq.~(\ref{eq:overall}), essentially reversing what we did to derive the differential version in Eq.~(\ref{eq:ugly}), we obtain the integrated equation of motion we have been seeking:
\begin{equation} \label{eq:final2p} \begin{split}
\SIG(t) \M[.5]=\M[.5] \SIG^1 \otimes \SIG^2 \,+\, \sin\!\big( \|\EMB\nu^1\| \|\EMB\nu^2\| \,t/4 \big)\, \big[\M[-0.25]\big[ \SIG^1 \otimes \SIG^2 ,\, \hat{\EMB\nu}^1 \otimes \hat{\EMB\nu}^2 \M[.1]\big]\M[-0.25]\big] \:- \M[3.2] &
\\[0.5ex] \cdots~ \big(1 - \cos\!\big( \|\EMB\nu^1\| \|\EMB\nu^2\| \,t/4 \big) \big)\, \big[\M[-0.25]\big[ \big[\M[-0.25]\big[ \SIG^1 \otimes \SIG^2 ,\, \hat{\EMB\nu}^1 \otimes \hat{\EMB\nu}^2 \M[.1]\big]\M[-0.25]\big] ,\, \hat{\EMB\nu}^1 \otimes \hat{\EMB\nu}^2 \M[.1]\big]\M[-0.25]\big] \,.
\end{split} \end{equation}
Equations (\ref{eq:form1a}--\ref{eq:form1b}) tell us (more or less) what the geometric interpretation of the two particle commutator is, so we turn our attention to the double commutator.
Geometric algebra shows that the double commutator of tensor products of the corresponding traceless $2\times2$ Hermitian matrices reduces to
\begin{multline}
2\, \big[ \big[ \MAT A \otimes \MAT C,\, \MAT B \otimes \MAT D \big],\, \MAT B \otimes \MAT D \big] \\
\begin{aligned}[t] =\M[.5]
& \big[ \HIP{\MAT A}{\MAT B} \big( \MAT P_{00} \otimes [ \MAT C, \MAT D ] \big) \,+\, \big( [ \MAT A, \MAT B ] \otimes \MAT P_{00} \big) \HIP{\MAT C}{\MAT D} \,,\: \MAT B \otimes \MAT D \big] \M[2] \\ =\M[.5]
& \HIP{\MAT A}{\MAT B}\, \MAT B \otimes [\. [\MAT C,\. \MAT D],\. \MAT D \.] \,+\, [\. [\MAT A,\. \MAT B],\. \MAT B \.] \otimes \MAT D\, \HIP{\MAT C}{\MAT D} ~,
\end{aligned} \end{multline}
so we will be done once we figure out what the double commutator of $2\times2$ matrices is.
On expanding the commutator and using the fact that for such matrices the anticommutator satisfies $\MAT{AB} + \MAT{BA} = \HIP{\MAT A}{\MAT B}\, \MAT P_{00\,}$, we get (including the real analogs):
\begin{equation} \label{eq:1pdc} \begin{split}
[\M[-0.12][\. [\M[-0.12][\, \MAT a ,\. \MAT b \,]\M[-0.12]] ,\. \MAT b \,]\M[-0.12]] \M[.5]\leftrightarrow\M[.5] &
[\. [\. \MAT A,\. \MAT B \.],\. \MAT B \.] \M[.5]=\M[.5] \MAT{AB}^2 \,-\, 2\, \MAT{BAB} \,+\, \MAT B^2 \MAT A \\ =\M[.5] &
\MAT A\. \| \MAT B \|^2 \,-\, 2\, (\MAT{BA} + \MAT{AB} - \MAT{AB})\. \MAT B \\ =\M[.5] &
2\, \big( \MAT A\. \| \MAT B \|^2 \,-\, \HIP{\MAT A}{\MAT B}\. \MAT B \big) \M[.5]\leftrightarrow\M[.5] \MAT a\, \|\. \MAT b \.\|^2 \,-\, \HIP{\MAT a}{\MAT b}\, \MAT b  ~.
\end{split} \end{equation}
Back-substitution of this and the corresponding expression for $[[\MAT C, \MAT D], \MAT D]$ into the preceding equation now yields:
\begin{multline}
\M[-1] \big[ \big[ \MAT A \otimes \MAT C,\, \MAT B \otimes \MAT D \big],\, \MAT B \otimes \MAT D \big] \\\begin{aligned}[t] &
\M[-.5]=\M[.5] \HIP{\MAT A}{\MAT B}\, \MAT B \otimes \big( \MAT C\, \|\MAT D\|^2 \,-\, \HIP{\MAT C}{\MAT D}\, \MAT D \big) \:+\: \big( \MAT A\, \|\MAT B\|^2 \,-\, \HIP{\MAT A}{\MAT B}\, \MAT B\big) \otimes \MAT D\, \HIP{\MAT C}{\MAT D} \\ &
\M[-.5]=\M[.5] \HIP{\MAT A}{\MAT B}\, \|\MAT D\|^2\, (\MAT B \otimes \MAT C) ~+~ \HIP{\MAT C}{\MAT D}\, \|\MAT B\|^2\, (\MAT A \otimes \MAT D) ~-~ 2\, \HIP{\MAT A}{\MAT B}\, \HIP{\MAT C}{\MAT D}\, (\MAT B \otimes \MAT D) .\M[-1]
\end{aligned} \end{multline}
Since no matrix products occur in this expression, we may transliterate directly to the corresponding real expression including the additional terms arising from setting $a_{00} = c_{00} = 1$ (cf.~Eq.~(\ref{eq:form1b})):
\begin{multline}
4\, \big[\M[-0.25]\big[ \big[\M[-0.25]\big[ (\MAT a + \MAT E_{00}) \otimes (\MAT c + \MAT E_{00}) ,\, \MAT b \otimes \MAT d \M[.1]\big]\M[-0.25]\big] ,\, \MAT b \otimes \MAT d \M[.1]\big]\M[-0.25]\big] \M[.5]= \\
\begin{aligned}[t]
\HIP{\MAT a}{\MAT b}\, \|\MAT{d}\|^2\, (\MAT b \otimes \MAT c) \:+\: \HIP{\MAT c}{\MAT d}\, \|\MAT b\|^2\, (\MAT a \otimes \MAT d)
~\cdots \M &
\\ -~ 2\, \HIP{\MAT a}{\MAT b}\, \HIP{\MAT c}{\MAT d}\, (\MAT b \otimes \MAT d) ~\cdots \M &
\\ +~ 2\, \big[\M[-0.25]\big[\, \MAT b \otimes [\M[-0.12][\. \MAT c ,\. \MAT d \. {]\M[-0.13]]} \,+\, [\M[-0.12][\. \MAT a ,\. \MAT b \.]\M[-0.12]] \otimes \MAT d ,\, \MAT b \otimes \MAT d \,\big]\M[-0.25]\big] \M[-1.5] & \M[1.5].
\end{aligned}
\end{multline}
The first of these additional terms (given on the last line of the above equation) may be evaluated via Eqs.~(\ref{eq:form1a}), (\ref{eq:form1b}) \& (\ref{eq:1pdc}) as indicated below,
\begin{equation} \begin{split}
\M[-1] 2\, \big[\M[-0.25]\big[\, \MAT b \otimes [\M[-0.12][\. \MAT c ,\. \MAT d \. {]\M[-0.15]]}  ,\, \MAT b \otimes \MAT d \,\big]\M[-0.25]\big] \M[.5]=\M[.75] &
{\|\. \MAT b \.\|}^2 \big( \MAT E_{00} \otimes [\M[-0.12][\. [\M[-0.12][\. \MAT c ,\. \MAT d \. ]\M[-0.13]] ,\. \MAT d \. ]\M[-0.13]] \big) \:+\: \big(\. [\M[-0.12][\. \MAT b ,\. \MAT b \. ]\M[-0.13]] \otimes \MAT E_{00} \big)\, \big\langle\. [\M[-0.12][\. \MAT c ,\. \MAT d \. ]\M[-0.13]] \.\big|\. \MAT d \. \big\rangle \\ =\M[.75] &
{\|\. \MAT b \.\|}^2\. \big(\, \MAT E_{00} \otimes (\. {\|\. \MAT d\. \|}^2 \. \MAT c \,-\, \HIP{\MAT c}{\MAT d}\, \MAT d \.) \big) ~,
\end{split} \end{equation}
with an analogous expression for the remaining term.
We now put in our previous values $\MAT a = \VP{\SIG}^1 \equiv \SIG^1 - \MAT E_{00}$, $\MAT b = \EMB\nu^1$, $\MAT c = \VP{\SIG}^2 \equiv \SIG^2 - \MAT E_{00}$ and $\MAT d = \EMB\nu^2$ and expand the result fully to get:
\begin{multline}
4\, \big[\M[-0.25]\big[ \big[\M[-0.25]\big[ \SIG^1 \otimes \SIG^2 ,\, \EMB\nu^1 \otimes \EMB\nu^2 \M[.1]\big]\M[-0.25]\big] ,\,\EMB\nu^1 \otimes \EMB\nu^2 \M[.1]\big]\M[-0.25]\big] \M[.5]= \\
\begin{aligned}[t]
\big\langle\, \SIG^1 \,\big|\, \EMB\nu^1 \big\rangle\, {\big\|\EMB\nu^2\big\|}^2 \, \big( \EMB\nu^1 \otimes \VP{\SIG}^2\. \big) \:+\: {\big\|\.\EMB\nu^1\big\|}^2\, \big\langle\, \SIG^2 \,\big|\, \EMB\nu^2\, \big\rangle\, \big( \VP{\SIG}^1 \otimes \EMB\nu^2\. \big) ~\cdots \M &
\\ -~ 2\, \big\langle\, \SIG^1 \,\big|\, \EMB\nu^1\, \big\rangle\, \big\langle\, \SIG^2 \,\big|\, \EMB\nu^2\, \big\rangle\, \big( \EMB\nu^1 \otimes \EMB\nu^2\. \big) ~\cdots \M &
\end{aligned} \\ \begin{aligned}[t]
+~{\big\|\. \EMB\nu^1\. \big\|}^2\, {\big\|\. \EMB\nu^2\. \big\|}^2\. \big( \MAT E_{00} \otimes \VP{\SIG}^2 \,+\, \VP{\SIG}^1 \otimes \MAT E_{00} \big) ~\cdots \M[1.25] &
\\ -~ \Big(  {\big\|\. \EMB\nu^1\. \big\|}^2\, \big\langle\, \SIG^2 \,\big|\, \EMB\nu^2\, \big\rangle\, \MAT E_{00} \otimes \EMB\nu^2 \:+\: \big\langle\, \SIG^1 \,\big|\, \EMB\nu^1\, \big\rangle\, {\big\|\. \EMB\nu^2\. \big\|}^2\, \EMB\nu^1 \otimes \MAT E_{00} \Big) . &
\end{aligned}
\end{multline}
Finally, we divide through by $\|\EMB\nu^1\|^2\, \|\EMB\nu^2\|^2$ to get the normalized ``vectors'' $\smash{\hat{\EMB\nu}}^1$ and $\smash{\hat{\EMB\nu}}^2$, replace the $\SIG$'s by $\VP{\SIG}$'s inside the traces (which doesn't change their values) and recombine terms to obtain:
\begin{multline}
4\, \big[\M[-0.25]\big[ \big[\M[-0.25]\big[ \SIG^1 \otimes \SIG^2 ,\, \hat{\EMB\nu}^1 \otimes \hat{\EMB\nu}^2 \M[.1]\big]\M[-0.25]\big] ,\, \hat{\EMB\nu}^1 \otimes \hat{\EMB\nu}^2 \M[.1]\big]\M[-0.25]\big] \M[.5]= \\
\begin{aligned}[t]
& \big( \VP{\SIG}^1 \:-\: \big\langle\, \VP\SIG^1 \,\big|\, \hat{\EMB\nu}^1\, \big\rangle\, \hat{\EMB\nu}^1 \big) \otimes \big( \big\langle\, \VP\SIG^2 \,\big|\, \hat{\EMB\nu}^2\, \big\rangle\, \hat{\EMB\nu}^2 \:+\: \MAT E_{00} \big) ~\cdots \M[2] \\ +~
& \big( \big\langle\, \VP\SIG^1 \,\big|\, \hat{\EMB\nu}^1\, \big\rangle\, \hat{\EMB\nu}^1 \:+\: \MAT E_{00} \big) \otimes \big( \VP{\SIG}^2 \:-\: \big\langle\, \VP\SIG^2 \,\big|\, \hat{\EMB\nu^2}\, \big\rangle\, \hat{\EMB\nu}^2 \big) ~.
\end{aligned}
\end{multline}
This has a fairly simple interpretation: The rejection of $\VP{\SIG}^1$ from $\hat{\EMB\nu}^1$ is tensored with the projection of $\VP{\SIG}^2$ onto $\hat{\EMB\nu}^2$ (plus the usual scalar part of $1$) and added to the projection of $\VP{\SIG}^1$ onto $\hat{\EMB\nu}^1$ (plus the scalar part) tensored with the rejection of $\VP{\SIG^2}$ from $\hat{\EMB\nu}^2$.
This should be contrasted with the single commutator (Eq.~(\ref{eq:form1a})), wherein the inner and outer products of each pair are tensored together both ways and added.

\section{Meta-Metamorphosis}
In the previous section we have shown how the Hamiltonians usually assumed for quantum computing with qubits can be integrated entirely within the real domain.
For a single qubit, the results could be interpreted as a simple Bloch vector rotation.
With a bi-axial interaction between two qubits, we also found that that the integrated expression had a reasonably nice geometric interpretation.
Algebraically, however, it is usually easier to integrate in the Hermitian domain, simply because Hermitian matrices are easier to diagonalize.
In this section, therefore, we shall derive formulae by which matrix representations of general superoperators can be translated from the Hermitian into the real domain, along with some specific examples of their utility.

Using the identity given in Eq.~(\ref{eq:oppr2prop}) together with our operator sum expressions for $\ALG U$ and its inverse (Eqs.~(\ref{eq:you}) \& (\ref{eq:uoy})), it is straightforward to show that an arbitrary superoperator $\ALG S$ with matrix representation $\EMB{\ALG S}$ acting on $\RHO$ via $\RHO \mapsto \EMB{\ALG S}\, \KET{\RHO}$ transforms into the real domain according to
\begin{equation} \label{eq:supopstfm}
\EMB{\ALG S} \M\stackrel{\ALG U}{\longleftrightarrow}\M 2^{-N\,} \EMB{\ALG Q} \bigg( \sum_{m'=0}^M\, \MAT P_{M,(M-m')} \otimes \MAT P_{m',0} \bigg) \EMB{\ALG S} \bigg( \sum_{m=0}^M\, \MAT P_{M,(M-m)} \otimes \MAT P_{m,0} \bigg) \OL{\EMB{\ALG Q}} ~,
\end{equation}
where $\EMB{\ALG Q} \equiv \DMAT(\.\KET{\MAT Q^{\otimes N}})$. The superoperator $\ALG S$ may be written in operator sum form versus the basis of elementary matrices \cite{Havel!QPT:03} as
\begin{equation}
\EMB{\ALG S}\, \KET{\RHO} ~=~ \bigg| \sum_{i,j=0}^{M,M}\, \sum_{k,\ell=0}^{M,M}\, s_{k\ell}^{ij}\, \MAT E_{ki\,} \RHO\M[0.05] \MAT E_{j\ell} \bigg\rangle ~=~ \sum_{i,j=0}^{M,M}\, \sum_{k,\ell=0}^{M,M}\, s_{k\ell}^{ij}\, \big( \MAT E_{\ell j} \otimes \MAT E_{ki} \big)\, \KET{\RHO} \,.
\end{equation}
On substituting the second of these equations (sans $\KET\RHO$) into the first and rearranging things a bit, the transformed superoperator becomes
\begin{equation}
2^{-N\,} \EMB{\ALG Q} \bigg( \sum_{i,j=0}^{M,M}\, \sum_{k,\ell=0}^{M,M}\, s_{k\ell}^{ij}\, \sum_{m,m'=0}^{M,M}\, \big( \MAT P_{M,(M-m')\,} \MAT E_{\X[1.3]\ell j\,} \MAT P_{M,(M-m)} \big) \otimes \big( \MAT P_{m',0\,} \MAT E_{ki\,} \MAT P_{m,0} \big)\! \bigg) \OL{\EMB{\ALG Q}} ~.
\end{equation}
Unfortunately, because each factor in the above Kronecker product depends on both indices $m$ and $m'$, this cannot be regarded as a transformation of the elementary matrix basis into the real domain.

Better insight can be obtained by looking at how the Choi matrix $\EMB{Choi}(\EMB{\ALG S})$ of the superoperator transforms.
This may be obtained from the propagating matrix $\EMB{\ALG S}$ simply by replacing Kronecker products of the elementary matrices by dyadic products of the corresponding columnized basis, but in the opposite order \cite{Havel!QPT:03}.
This task is facilitated by expressing the left- and right-multiplication by diagonal matrices as a Hadamard product, using the well-known formula \cite{HaShViCo:01}
\begin{equation} \label{eq:wellknown}
\DMAT(\MAT a)\, \MAT X\, \DMAT^\dag(\MAT b) ~=~ (\MAT{ab}^\dag) \odot \MAT X ~,
\end{equation}
which is essentially a special case of Eq.~(\ref{eq:form2}).
This allows the transformed superoperator to be rewritten as
\begin{multline}
2^{-N\,} \Big( \big|\, \MAT Q^{\otimes N} \big\rangle\, \big\langle\. \MAT Q^{\otimes N} \big| \Big) \odot \sum_{i,j=0}^{M,M}\, \sum_{k,\ell=0}^{M,M}\, s_{k\ell}^{ij} ~\cdots \\
\cdots \sum_{m,m'=0}^{M,M}\, \Big(\! \big( \MAT P_{M,(M-m')\,} \MAT E_{\X[1.3]\ell j\,} \MAT P_{M,(M-m)} \big) \otimes \big( \MAT P_{m',0\,} \MAT E_{ki\,} \MAT P_{m,0} \big)\! \Big) ~.
\end{multline}
The advantage of this form is that the Hadamard product commutes with the $\EMB{Choi}$ operator (since it rearranges the entries of the product's operands identically), giving us
\begin{align}
\M[-0.25] \EMB{Choi}(\EMB{\ALG S}) \M[0.25]\stackrel{\ALG U}{\longleftrightarrow}\M[0.5] &
2^{-N\,} \EMB{Choi}\Big( \big|\, \MAT Q^{\otimes N} \big\rangle\, \big\langle\. \MAT Q^{\otimes N} \big| \Big) \odot \sum_{i,j=0}^{M,M}\, \sum_{k,\ell=0}^{M,M}\, s_{k\ell}^{ij} ~\cdots \notag\\ &
\cdots\, \sum_{m,m'=0}^{M,M}\, \EMB{Choi}\Big(\! \big( \MAT P_{M,(M-m')\,} \MAT E_{\X[1.3]\ell j\,} \MAT P_{M,(M-m)} \big) \otimes \big( \MAT P_{m',0\,} \MAT E_{ki\,} \MAT P_{m,0} \big)\! \Big) \notag\\ =\M[0.5] &
2^{-N\,} \big( \OL{\MAT Q}^{\.\otimes N\!} \otimes \MAT Q^{\otimes N} \big) \odot \sum_{i,j=0}^{M,M}\, \sum_{k,\ell=0}^{M,M}\, s_{k\ell}^{ij} ~\cdots \notag\\ &
\cdots\, \sum_{m,m'=0}^{M,M}\, \Big( \big|\, \MAT P_{m',0\,} \MAT E_{ki\,} \MAT P_{m,0} \big\rangle \big\langle \MAT P_{M,(M-m')\,} \MAT E_{\X[1.3]\ell j\,} \MAT P_{M,(M-m)} \big| \Big) \notag\\ =\M[0.5] &
2^{-N\,} \big( \OL{\MAT Q}^{\.\otimes N\!} \otimes \MAT Q^{\otimes N} \big) \odot \sum_{i,j=0}^{M,M}\, \sum_{k,\ell=0}^{M,M}\, s_{k\ell}^{ij}\sum_{m,m'=0}^{M,M}\, \big( \MAT P_{m,0} \otimes \MAT P_{m',0} \big)\, \KET{\MAT E_{ki\,}} ~\cdots\notag\\ &
\cdots~ \big\langle\, \MAT E_{\X[1.3]\ell j\,} \big| \big( \MAT P_{M,(M-m)} \otimes \MAT P_{M,(M-m')} \big) \\\notag =\M[0.5] &
2^{-N\,} \big( \OL{\MAT Q}^{\.\otimes N\!} \otimes \MAT Q^{\otimes N} \big) \odot \sum_{m,m'=0}^{M,M}\, \big( \MAT P_{m,0} \otimes \MAT P_{m',0} \big)\, \sum_{i,j=0}^{M,M}\, \sum_{k,\ell=0}^{M,M}\, s_{k\ell}^{ij} ~\cdots  \\\notag &
\cdots~ \big( \MAT E_{ij} \otimes \MAT E_{k\ell} \big)\big( \MAT P_{M,(M-m)} \otimes \MAT P_{M,(M-m')} \big) \\\notag \equiv\M[0.5] &
2^{-N\,} \big( \OL{\MAT Q}^{\.\otimes N\!} \otimes \MAT Q^{\otimes N} \big) \odot \sum_{m=0}^{M'}\, \MAT P_{m,0}\; \EMB{Choi}(\EMB{\ALG S})\; \MAT P_{M',(M'-m)} ~,
\end{align}
where $M' \equiv (M+1)^2 - 1 = 2^{2N} - 1$ and we have used the relation $\KET{\MAT E_{ki}}\, \BRA{\MAT E_{\ell j}} = (\MAT e_i \otimes \MAT e_k)(\MAT e_j \otimes \MAT e_\ell)^\top = \MAT E_{ij} \otimes \MAT E_{k\ell}$.

In other words, the Choi matrix of a superoperator maps into the real domain much like a density matrix on twice as many qubits.
In fact we can write the \emph{real} transformation matrix $\EMB{\ALG T}$, which acts on the real density matrix as $\EMB{\ALG T}\, \KET{\SIG}$, in the following compact form:
\begin{equation}
\EMB{\ALG T} \M[0.7]=\M[0.7] 2^{-N}\, \EMB{Choi}\Big(\, \ALG U\big(\, \EMB{Choi}( \EMB{\ALG S} )\, \big) \Big)\, \OL{\EMB{\ALG Q}}^{\,2} ~.
\end{equation}
Turning this around, we also find that we can express the Choi matrix of $\ALG S$ in the Pauli basis as
\begin{equation} \begin{split}
& 2^{-N}\; \EMB{Choi}(\EMB{\ALG S}) \M[0.7]=\M[0.7] \ALG U^{-1}\big( \EMB{Choi}( \EMB{\ALG T}\. \EMB{\ALG Q}^2 \.) \big) \\ \equiv\M &
\sum_{i,j=0}^{M,M} \sum_{k,\ell=0}^{M,M}\, \ALG U^{-1}\big(\. t_{k\ell}^{ij}\; \EMB{Choi}\big( (\MAT E_{\ell j} \otimes \MAT E_{ki})\. \EMB{\ALG Q}^2 \.\big) \big) \\ =\M &
\sum_{i,j=0}^{M,M} \sum_{k,\ell=0}^{M,M}\, t_{k\ell}^{ij}~ \ALG U^{-1}\Big( \EMB{Choi}(\MAT E_{\ell j} \otimes \MAT E_{ki}) \odot \EMB{Choi}\big(\, \KET{\MAT 1\, \MAT 1^\top} \BRA{\MAT Q^{\otimes N\!} \odot \MAT Q^{\otimes N}}\, \big)\! \Big) \\ =\M &
\sum_{i,j=0}^{M,M} \sum_{k,\ell=0}^{M,M}\, t_{k\ell}^{ij}\; \ALG U^{-1}\big( \MAT E_{ij} \otimes \MAT E_{k\ell} \big) \odot \big( \big( \MAT Q^{\otimes N\!} \odot \MAT Q^{\otimes N} \big) \otimes \big( \MAT 1\. \MAT 1^\top \big) \big) \big) \\ =\M &
2^{-2N}\, \big( \big( \MAT Q^{\otimes N\!} \odot \MAT Q^{\otimes N} \big) \otimes \big( \MAT 1\. \MAT 1^\top \big) \big) \odot \sum_{i,j=0}^{M,M} \sum_{k,\ell=0}^{M,M}\, t_{k\ell}^{ij}\; \big( \MAT P_{ij} \otimes \MAT P_{k\ell} \big) ~,
\end{split} \end{equation}
where $\MAT 1$ denotes a column vector of ones of the appropriate size.

It is time for our examples!
We shall begin with operator sums for single qubit rotations, and go on to show how rotations about the $\SIG[3]$ axis as well as the $\SIG[3]^1\SIG[3]^2$ interaction between two qubits can be compactly described in the real domain using Hadamard products.
We close by showing that this description also extends quite nicely to $\SIG[3]$ dephasing as well as nonunital relaxation back towards a nonrandom equilibrium state (i.e.~$T_2$ and $T_1$ relaxation in NMR parlance).

Consider the Choi matrix of the propagator which rotates a single qubit (Eqs.~(\ref{eq:form0}--\ref{eq:r2}) above):
\begin{align}
& \EMB{Choi}\big( \MAT P_{00} \otimes \MAT P_{00} \,+\, \sin(\|\EMB\nu\|\M[.1]t/2\M[.1])\, \EMB{\ALG R}_{\widehat{\EMB\nu}} \,+\, (1-\cos(\|\EMB\nu\|\M[.1]t/2\M[.1]))\, \EMB{\ALG R}_{\widehat{\EMB\nu}}^{\,2} \big) \notag\\ =\M[0.5] &
\begin{bmatrix} ~1~&~0~&~0~&~1~\\ ~0~&~0~&~0~&~0~\\  ~0~&~0~&~0~&~0~\\ ~1~&~0~&~0~&~1~ \end{bmatrix} ~+~ \sin(\|\EMB\nu\|\M[.1]t/2\M[.1]) \begin{bmatrix} ~0&0&0&0\\ ~0&0&-\hat\nu_{11}&\hat\nu_{10}\\ ~0&\hat\nu_{11}&0&-\hat\nu_{10}\\ ~0&-\hat\nu_{01}&\hat\nu_{01}&0 \end{bmatrix} ~+~ \cdots \\\notag &
\cdots~ (1-\cos(\|\EMB\nu\|\M[.1]t/2\M[.1])) \begin{bmatrix} ~0&0&0& -\hat\nu_{10}^2-\hat\nu_{11}^2 \\ ~0&0& \hat\nu_{10}\hat\nu_{01} & \hat\nu_{01}\hat\nu_{11} \\ ~0& \hat\nu_{10}\hat\nu_{01} &0& \hat\nu_{01}\hat\nu_{11} \\ -\hat\nu_{01}^2-\hat\nu_{11}^2 & \hat\nu_{10}\hat\nu_{11} & \hat\nu_{10}\hat\nu_{11} & -\hat\nu_{01}^2-\hat\nu_{10}^2 \end{bmatrix} ~,
\end{align}
where the ``hat'' on the $\nu$'s indicates normalization by $\|\EMB\nu\|$.
A general operator sum can be derived from this matrix (as well as by expanding the implied commutators via Eq.~(\ref{eq:1pc})), but since this is a bit involved we shall restrict ourselves to rotations by an angle $\vartheta =\|\EMB\nu\|\M[0.1]t/2$ about the $\LAB x$, $\LAB y$ or $\LAB z$ coordinate axes.
On substituting $\hat\nu_{10} = 1$ and $\hat\nu_{01} = \hat\nu_{11} = 0$ into the above Choi matrix we obtain the following singular value decomposition:
\begin{equation} \begin{split}
\begin{bmatrix} ~1~&~0~&~0~& \cos(\vartheta)\\ ~0~&~0~&~0~& \sin(\vartheta)\\ ~0~&~0~&~0~& -\sin(\vartheta)\\ ~1~&~0~&~0~& \cos(\vartheta) \end{bmatrix} \M[0.5]=\M[0.5] &
\begin{bmatrix} \cos(\vartheta/2)&\sin(\vartheta/2)\\ \sin(\vartheta/2)~&-\cos(\vartheta/2)\\ -\sin(\vartheta/2)~&~\cos(\vartheta/2)\\ \cos(\vartheta/2)&\sin(\vartheta/2) \end{bmatrix} \cdots \\ &
\cdots \begin{bmatrix} \cos(\vartheta/2)&0\\ 0&\sin(\vartheta/2) \end{bmatrix}\M[-0.25] \begin{bmatrix} ~1~&~0~&~0~&~1~\\ ~1~&~0~&~0~&-1~ \end{bmatrix} ~,
\end{split} \end{equation}
as may be readily verified using the usual half-angle formulae.
The corresponding operator sum for rotation by $\vartheta$ about the $\LAB x$-axis is simply:
\begin{equation} \begin{split}
\ALG U\Big( \EMB e^{-\imath(\vartheta/2)\MAT P_{10}} \RHO\, \EMB e^{~\imath(\vartheta/2)\MAT P_{10}} \Big) \M[0.5]=\M[0.5] &
\cos(\vartheta/2) \begin{bmatrix} \cos(\vartheta/2)&-\sin(\vartheta/2)\\ \sin(\vartheta/2)&\cos(\vartheta/2) \end{bmatrix}\M[-0.5] \begin{bmatrix} \text{\small$1$}&\sigma_{01}\\ \sigma_{10}&\sigma_{11} \end{bmatrix} ~+~ \cdots \\ &
\sin(\vartheta/2) \begin{bmatrix} \sin(\vartheta/2)&\cos(\vartheta/2)\\ -\cos(\vartheta/2)&\sin(\vartheta/2) \end{bmatrix}\M[-0.5] \begin{bmatrix} \text{\small$1$}&\sigma_{01}\\ \sigma_{10}&\sigma_{11} \end{bmatrix}\M[-0.5] \begin{bmatrix} ~1~&~0~\\ ~0~&-1~ \end{bmatrix} .
\end{split} \end{equation}
In a similar fashion, it can be shown that the operator sum for a $\LAB y$-rotation is:
\begin{equation} \begin{split}
\ALG U\Big( \EMB e^{-\imath(\vartheta/2)\MAT P_{01}} \RHO\, \EMB e^{~\imath(\vartheta/2)\MAT P_{01}} \Big) \M[0.5]=\M[0.5] &
\cos(\vartheta/2) \begin{bmatrix} \text{\small$1$}&\sigma_{01}\\ \sigma_{10}&\sigma_{11} \end{bmatrix}\M[-0.5] \begin{bmatrix} \cos(\vartheta/2)&-\sin(\vartheta/2)\\ \sin(\vartheta/2)&\cos(\vartheta/2) \end{bmatrix} ~+~ \cdots \\ &
\sin(\vartheta/2) \begin{bmatrix} ~1~&~0~\\ ~0~&-1~ \end{bmatrix}\M[-0.5] \begin{bmatrix} \text{\small$1$}&\sigma_{01}\\ \sigma_{10}&\sigma_{11} \end{bmatrix}\M[-0.5] \begin{bmatrix} \sin(\vartheta/2)&\cos(\vartheta/2)\\ -\cos(\vartheta/2)&\sin(\vartheta/2) \end{bmatrix} .
\end{split} \end{equation}

For a $\LAB z$-rotation, on the other hand, the Choi matrix turns out to be rank $4$ with singular value decomposition: 
\begin{equation} \begin{split} &
\begin{bmatrix} 1&0&0&\cos(\vartheta)\\ 0&0&-\sin(\vartheta)&0\\ 0&\sin(\vartheta)&0&0\\ \cos(\vartheta)&0&0&1 \end{bmatrix} \M[0.5]=\M[0.5]
\begin{bmatrix} ~1~&~0~&~0~&~1~\\ ~0~&~1~&-1~&~0~\\ ~0~&~1~&~1~&~0~\\ ~1~&~0~&~0~&-1~ \end{bmatrix} \cdots \\ &
\cdots \begin{bmatrix} \cos^2(\vartheta/2)&0&0&0\\ 0&\cos(\vartheta/2)\sin(\vartheta/2)&0&0\\ 0&0&\cos(\vartheta/2)\sin(\vartheta/2)&0\\ 0&0&0&\sin^2(\vartheta/2) \end{bmatrix}\M[-0.5] \begin{bmatrix} ~1~&~0~&~0~&~1~\\ ~0~&1~&-1~&~0~\\ ~0~&~1~&~1~&~0~\\ ~1~&~0~&~0~&-1~ \end{bmatrix} .
\end{split} \end{equation}
This corresponds to the operator sum
\begin{equation} \begin{split}
\ALG U\Big( \EMB e^{-\imath(\vartheta/2)\MAT P_{11}} \RHO\, \EMB e^{~\imath(\vartheta/2)\MAT P_{11}} \Big) \M[0.5]=\M[0.5] &
\cos^2(\vartheta/2)\, \SIG ~+~ \sin^2(\vartheta/2)\, \MAT P_{11\,} \SIG\, \MAT P_{11} ~+~ \cdots \\ &
\cdots~ \imath \cos(\vartheta/2)\sin(\vartheta/2) \big( \MAT P_{10\,} \SIG\, \MAT P_{01} ~+~ \MAT P_{01\,} \SIG\, \MAT P_{10} \big) ~,
\end{split} \end{equation}
which has the pleasant feature that the trigonometric functions occur as scalar factors in each term and not embedded in the operators.
This enables us to use Eq.~(\ref{eq:wellknown}) to rewrite it in terms of the Hadamard product as follows:
\begin{equation} \begin{split}
\ALG U\Big( \EMB e^{-\imath(\vartheta/2)\MAT P_{11}} \RHO\, \EMB e^{~\imath(\vartheta/2)\MAT P_{11}} \Big) \M[0.5] &
=\M[0.5]\begin{bmatrix} 1&\cos(\vartheta)\\ \cos(\vartheta)&1 \end{bmatrix} \odot \begin{bmatrix} \text{\small$1$}&\sigma_{01}\\ \sigma_{10}&\sigma_{11} \end{bmatrix} ~+~ \cdots \\ \cdots~ \begin{bmatrix} ~0~&~1~\\ ~1~&~0~ \end{bmatrix}\M[-0.25] &
\left( \begin{bmatrix} 0&-\sin(\vartheta)\\ \sin(\vartheta)&0 \end{bmatrix} \odot \begin{bmatrix} \text{\small$1$}&\sigma_{01}\\ \sigma_{10}&\sigma_{11} \end{bmatrix} \right)\M[-0.25] \begin{bmatrix} ~0~&~1~\\ ~1~&~0~ \end{bmatrix} ~.
\end{split} \end{equation}

As our next example, wherein the Hadamard product enables even greater simplifications, consider an ``Ising-type'' interaction between two qubits of the form $\SIG[3] \otimes \SIG[3] = \MAT P_{33}$, which is also known as ``weak scalar coupling'' in NMR \cite{ErnBodWok:87}.
An operator sum expression for this could be obtained by expanding the general formula given in Eq.~(\ref{eq:final2p}), but because this Hamiltonian is again diagonal in the $\SIG[3]$ eigenbasis a simpler expression can be obtained directly starting from the diagonal matrix of the corresponding propagator, i.e.
\begin{equation} \begin{split}
\big|\. \EMB e^{-\imath \MAT P_{33} \pi J t / 2} \RHO\M[0.1] \EMB e^{\,\imath \MAT P_{33} \pi J t / 2}\M[0.1] \big\rangle \M[0.5]=\M[0.5] &
\big( \EMB e^{\,\imath \MAT P_{33} \pi J t / 2} \otimes \EMB e^{-\imath \MAT P_{33} \pi J t / 2} \big)\M[0.1] \KET{\RHO} \\ =\M[0.5] &
\DMAT\big(\M[0.1] \KET{\MAT J(t)} \big) \KET{\RHO} \M[0.5]\equiv\M[0.5] \EMB{\ALG J}(t)\, \KET{\RHO} ~,
\end{split} \end{equation}
where
\begin{equation}
\MAT J(t) ~\equiv~ \begin{bmatrix} 1 & e^{\,\imath \pi J t} & e^{\,\imath \pi J t} & 1 \\ e^{-\imath \pi J t} & 1 & 1 & e^{-\imath \pi J t} \\ e^{-\imath \pi J t} & 1 & 1 & e^{-\imath \pi J t} \\ 1 & e^{\,\imath \pi J t} & e^{\,\imath \pi J t} & 1 \end{bmatrix} .
\end{equation}
On transforming this into the real domain via Eq.~(\ref{eq:supopstfm}), we find that $\EMB{\ALG J}$ has been converted into the sum of the diagonal matrix
\begin{equation}
\EMB{\ALG C}(t) ~\equiv~ \DMAT\M[-0.25]\left(\M[0.25] \begin{picture}(1,50)(0,50) \thicklines \put(0,10){\line(0,1){84}} \end{picture} 
\begin{bmatrix} 1 & \cos( \pi J t) & \cos( \pi J t) & 1 \\ \cos( \pi J t) & 1 & 1 & \cos( \pi J t) \\ \cos( \pi J t) & 1 & 1 & \cos( \pi J t) \\ 1 & \cos( \pi J t) & \cos( \pi J t) & 1 \end{bmatrix} 
\begin{picture}(10,50)(0,39) \thicklines \put(0,0){\line(1,6){7}} \put(0,84){\line(1,-6){7}} \end{picture} \M[-0.25]\right) ~,
\end{equation}
and the anti-diagonal matrix
\begin{equation}
\MAT P_{15,0}\M[0.25] \EMB{\ALG S}(t)\M[0.1] ~\equiv~ \DMAT\M[-0.25]\left(\M[0.25] \begin{picture}(1,50)(0,50) \thicklines \put(0,10){\line(0,1){84}} \end{picture} 
\begin{bmatrix} 0 & \sin( \pi J t) & \sin( \pi J t) & 0 \\ \sin( \pi J t) & 0 & 0 & -\sin( \pi J t) \\ \sin( \pi J t) & 0 & 0 & -\sin( \pi J t) \\ 0 & -\sin( \pi J t) & -\sin( \pi J t) & 0 \end{bmatrix}  
\begin{picture}(10,50)(0,39) \thicklines \put(0,0){\line(1,6){7}} \put(0,84){\line(1,-6){7}} \end{picture} \M[-0.25]\right) ~,
\end{equation}
where the (self-inverse) left factor of $\MAT P_{15,0} = \SIG[1]^{\otimes4}$ simply reverses the order of the rows.
The nonzero entries of the Choi matrix of $\EMB{\ALG C}(t) + \EMB{\ALG S}(t)$ turn out to comprise two $4\times4$ blocks along the diagonal, which are exactly the two matrices above, i.e.
\begin{equation} \begin{split}
\EMB{\ALG C}(t) ~=~ & \DMAT\big( \big| \EMB{\ALG E}_{\EMB{\ALG C}\,} \EMB{Choi}( \EMB{\ALG C}(t) ) \EMB{\ALG E}_{\EMB{\ALG C}\,}^\top \big\rangle \big) \\ \text{and}\quad
\MAT P_{15,0}\M[0.25] \EMB{\ALG S}(t) ~=~ & \DMAT\big( \big| \EMB{\ALG E}_{\EMB{\ALG S}\,} \EMB{Choi}( \EMB{\ALG S}(t) ) \EMB{\ALG E}_{\EMB{\ALG S}\,}^\top \big\rangle \big)\M[0.1] ~,
\end{split} \end{equation}
wherein
\begin{equation}
\EMB{\ALG E}_{\EMB{\ALG C}} ~\equiv~ \sum_{i=0}^3 \MAT e_i (\MAT e_i\M[0.1] \otimes \MAT e_i)^\top \qquad\text{and}\qquad \EMB{\ALG E}_{\EMB{\ALG S}} ~\equiv~ \sum_{i=0}^3 \MAT e_i\M[0.1] (\MAT e_i \otimes \MAT e_{3-i})^\top
\end{equation}
project out the rows / columns of their respective blocks.

Thus we can obtain the desired operator sum representation by computing the eigenvalues and eigenvectors of the $4\times4$ symmetric matrices
\begin{equation}
\MAT C(t) ~\equiv~ \EMB{\ALG E}_{\EMB{\ALG C}\,} \EMB{Choi}( \EMB{\ALG C}(t) ) \EMB{\ALG E}_{\EMB{\ALG C}\,}^\top \quad\text{and}\quad \MAT S(t) ~\equiv~ \EMB{\ALG E}_{\EMB{\ALG S}\,} \EMB{Choi}( \EMB{\ALG S}(t) ) \EMB{\ALG E}_{\EMB{\ALG S}\,}^\top ~,
\end{equation}
letting the operators' matrices be the diagonal / anti-diagonal matrices formed from the entries of these eigenvectors, and multiplying each term in the sum by the corresponding eigenvalue.
The results are
\begin{equation}
\EMB{\ALG C}(t)\, \KET{\SIG} \M[0.5]=\M[0.5] \Big|\, \HALF\, \big( 1 + \cos(\pi J t) \big)\, \SIG ~+~ \HALF\, \big( 1 - \cos(\pi J t) \big)\, \MAT P_{33}\M[0.2] \SIG\M[0.3] \MAT P_{33} \,\Big\rangle
\end{equation}
and
\begin{equation} \begin{split}
\! \EMB{\ALG S}(t)\, \KET{\SIG} \M[0.5]=\M[0.5] \HALF\, \sin(\pi J t)\, \Big|\, \MAT P_{30}\M[0.25] \DMAT\big( [\M[0.3]1,\,1,\,1,-1\M[0.1]] \big)\M[0.2] \SIG\M[0.3] \DMAT\big( [\M[0.3]1,\,1,\,1,-1\M[0.1]] \big)\M[0.1] \MAT P_{30} &
~\cdots \\ \cdots~ -~ \MAT P_{30}\M[0.25] \DMAT\big( [\M[0.1]-1,\,1,\,1,\,1\M[0.1]] \big)\M[0.2] \SIG\M[0.3] \DMAT\big( [\M[0.1]-1,\,1,\,1,\,1\M[0.1]] \big)\M[0.1] \MAT P_{30}\Big\rangle & ~.
\end{split} \end{equation}
By using Eq.~(\ref{eq:wellknown}) to replace these operator sums by Hadamard products and taking advantage of the symmetry of $\MAT S(t)$, however, we can obtain an even simpler expression, namely
\begin{equation} \label{eq:simpler}
\big( \EMB{\ALG C}(t) + \EMB{\ALG S}(t) \big)\M[0.1] \KET{\SIG} ~=~ \big| \MAT C(t) \odot \SIG \,-\, \MAT S(t) \odot (\MAT P_{30}\M[0.15] \SIG\M[0.15] \MAT P_{30}) \big\rangle ~.
\end{equation}

Finally, we show how one can also use Hadamard products with the real density matrix to describe simple relaxation processes, in a manner similar to that described in \citet{HaShViCo:01} for the usual Hermitian density matrix.
For a single qubit undergoing $T_1$ (dissipation) and $T_2$ (decoherence) relaxation, the time derivative is given by:
\begin{equation}
\partial_{t\,} \SIG(t) ~=~ -\MAT R \odot \SIG(t) ~\equiv~ -\begin{bmatrix} 0&1/T_2\\ 1/T_2&1/T_1 \end{bmatrix} \odot \begin{bmatrix} \text{\small$1$}&\sigma_{01}(t)\\ \sigma_{10}(t)&\sigma_{11}(t) \end{bmatrix} ~.
\end{equation}
Assuming that these relaxation processes are uncorrelated, this can immediately be extended to any number of qubits using the fact that the Hadamard product satisfies the mixed product formula with the Kronecker product (Eq.~(\ref{eq:mixed})).
In the case of two qubits relaxing with Hadamard relaxation matrices $\MAT R^1$, $\MAT R^2$, for example, we obtain
\begin{equation} \begin{split}
\partial_{t\,} \SIG(t) \M[0.5]=\M[0.5] & -\! \big( \MAT R^1 \otimes (\MAT{11}^\top) + (\MAT{11}^\top) \otimes \MAT R^2 \big) \odot \SIG(t) \\ =\M[0.5] &
- \begin{bmatrix} 0&1/T_2^2& 1/T_2^1&1/T_2^1+1/T_2^2\\  1/T_2^2&1/T_1^2& 1/T_2^1+1/T_2^2&1/T_2^1+1/T_1^2\\ 1/T_2^1&1/T_2^1+1/T_2^2&1/T_1^1&1/T_1^1+1/T_2^2\\ 1/T_2^1+1/T_2^2&1/T_2^1+1/T_1^2&1/T_1^1+1/T_2^2&1/T_1^1+1/T_1^2 \end{bmatrix} \odot\, \SIG(t) ~.
\end{split} \end{equation}
where $\MAT 1$ is a $4\times1$ vector of $1$'s.
The fact that uncorrelated $T_1$ as well as $T_2$ relaxation can be extended so easily to multiple spins in this way is actually a significant advantage of the real density matrix over the Hermitian, since in the latter case the diagonal terms are mixtures of terms decaying at differing rates, substantially complicating their treatment via Hadamard products \cite{HaShViCo:01}.

When correlations are present, however, these advantages are largely lost, since then the off-diagonal entries of the real density matrix consist of mixtures of terms with differing  decay rates (fortunately, $T_1$ relaxation is usually largely uncorrelated \cite{ErnBodWok:87}).
Let us work through the case of two qubits in detail, assuming for simplicity that the $T_2$ relaxation processes at the two qubits are totally correlated and have the same rate $1/T_2$, as for example in an NMR gradient-diffusion experiment \cite{HaShViCo:01}.
In this case the Hadamard relaxation matrix for the Hermitian density matrix has the form
\begin{equation}
\MAT R ~=~ \frac1{T_2} \begin{bmatrix} ~0~&~1~&~1~&~4~\\ ~1~&~0~&~0~&~1~\\ ~1~&~0~&~0~&~1~\\  ~4~&~1~&~1~&~0~ \end{bmatrix} ~,
\end{equation}
and the corresponding $16\times16$ diagonal relaxation superoperator
\begin{equation}
\EMB{\ALG R} ~=~ \DMAT\big(\. \KET{\MAT R} \big)
\end{equation}
is easily exponentiated into a diagonal matrix of survival probabilities for the entries of the (traceless part of the) Hermitian density matrix.
In this case, however, it turns out to be almost as easy, but more revealing, to convert $\EMB{\ALG R}$ into the real domain and perform the integration there.
The result of the first step is
\begin{equation}
\EMB{\ALG R} ~\stackrel{\ALG U}{\longleftrightarrow}~ \frac1{T_2}
\left[ \begin{smallmatrix}
0\:&\:0\:&\:0\:&\:0\:&\:0\:&\:0\:&\:0\:&\:0\:&\:0\:&\:0\:&\:0\:&\:0\:&\:0\:&\:0\:&\:0\:&\:0\\[0.25ex]
0\:&\:1\:&\:0\:&\:0\:&\:0\:&\:0\:&\:0\:&\:0\:&\:0\:&\:0\:&\:0\:&\:0\:&\:0\:&\:0\:&\:0\:&\:0\\[0.25ex]
0\:&\:0\:&\:1\:&\:0\:&\:0\:&\:0\:&\:0\:&\:0\:&\:0\:&\:0\:&\:0\:&\:0\:&\:0\:&\:0\:&\:0\:&\:0\\[0.25ex]
0\:&\:0\:&\:0\:&\:2\:&\:0\:&\:0\:&\:0\:&\:0\:&\:0\:&\:0\:&\:0\:&\:0\:&\!-2\:&\:0\:&\:0\:&\:0\\[0.25ex]
0\:&\:0\:&\:0\:&\:0\:&\:1\:&\:0\:&\:0\:&\:0\:&\:0\:&\:0\:&\:0\:&\:0\:&\:0\:&\:0\:&\:0\:&\:0\\[0.25ex]
0\:&\:0\:&\:0\:&\:0\:&\:0\:&\:0\:&\:0\:&\:0\:&\:0\:&\:0\:&\:0\:&\:0\:&\:0\:&\:0\:&\:0\:&\:0\\[0.25ex]
0\:&\:0\:&\:0\:&\:0\:&\:0\:&\:0\:&\:2\:&\:0\:&\:0\:&\:2\:&\:0\:&\:0\:&\:0\:&\:0\:&\:0\:&\:0\\[0.25ex]
0\:&\:0\:&\:0\:&\:0\:&\:0\:&\:0\:&\:0\:&\:1\:&\:0\:&\:0\:&\:0\:&\:0\:&\:0\:&\:0\:&\:0\:&\:0\\[0.25ex]
0\:&\:0\:&\:0\:&\:0\:&\:0\:&\:0\:&\:0\:&\:0\:&\:1\:&\:0\:&\:0\:&\:0\:&\:0\:&\:0\:&\:0\:&\:0\\[0.25ex]
0\:&\:0\:&\:0\:&\:0\:&\:0\:&\:0\:&\:2\:&\:0\:&\:0\:&\:2\:&\:0\:&\:0\:&\:0\:&\:0\:&\:0\:&\:0\\[0.25ex]
0\:&\:0\:&\:0\:&\:0\:&\:0\:&\:0\:&\:0\:&\:0\:&\:0\:&\:0\:&\:0\:&\:0\:&\:0\:&\:0\:&\:0\:&\:0\\[0.25ex]
0\:&\:0\:&\:0\:&\:0\:&\:0\:&\:0\:&\:0\:&\:0\:&\:0\:&\:0\:&\:0\:&\:1\:&\:0\:&\:0\:&\:0\:&\:0\\[0.25ex]
0\:&\:0\:&\:0\:&\!-2\:&\:0\:&\:0\:&\:0\:&\:0\:&\:0\:&\:0\:&\:0\:&\:0\:&\:2\:&\:0\:&\:0\:&\:0\\[0.25ex]
0\:&\:0\:&\:0\:&\:0\:&\:0\:&\:0\:&\:0\:&\:0\:&\:0\:&\:0\:&\:0\:&\:0\:&\:0\:&\:1\:&\:0\:&\:0\\[0.25ex]
0\:&\:0\:&\:0\:&\:0\:&\:0\:&\:0\:&\:0\:&\:0\:&\:0\:&\:0\:&\:0\:&\:0\:&\:0\:&\:0\:&\:1\:&\:0\\[0.25ex]
0\:&\:0\:&\:0\:&\:0\:&\:0\:&\:0\:&\:0\:&\:0\:&\:0\:&\:0\:&\:0\:&\:0\:&\:0\:&\:0\:&\:0\:&\:0
\end{smallmatrix} \right] ~,
\end{equation}
which again has nonzero entries only on its diagonal and its anti-diagonal.
Representing the action of this matrix on $\KET{\SIG}$ as a sum of a Hadamard product and a Hadamard product coupled with row/column inversion as in Eq.~(\ref{eq:simpler}), we obtain the real equation of motion:
\begin{equation}
-T_2\, \partial_t\, {\SIG}(t) \M[.5]=\M[.3] \begin{bmatrix} ~0~&~1~&~1~&~2~\\ ~1~&~0~&~2~&~1~\\ ~1~&~2~&~0~&~1~\\ ~2~&~1~&~1~&~0~ \end{bmatrix} \!\odot\. \SIG(t) ~+~ \begin{bmatrix} ~0~&~0~&~0~&-2~\\ ~0~&~0~&~2~&~0~\\ ~0~&~2~&~0~&~0~\\ -2~&~0~&~0~&~0~ \end{bmatrix} \odot\, \big( \MAT P_{30}\M[0.1] \SIG(t) \M[0.1]\MAT P_{30} \big)
\end{equation}
It may readily be verified that the operations on $\SIG$ which occur in the two terms of this expression commute, and hence this equation can be integrated by exponentiating them separately.
With the first term this leads to a simple Hadamard (entrywise) exponential \cite{HaShViCo:01}, namely
\begin{equation}
\MAT D(t) ~\equiv~ \MAT{Exp}_{\.\odot\!} \left( -\frac{t}{T_2} \begin{bmatrix} ~0~&~1~&~1~&~2~\\ ~1~&~0~&~2~&~1~\\ ~1~&~2~&~0~&~1~\\ ~2~&~1~&~1~&~0~ \end{bmatrix} \right) ~=~ \begin{bmatrix} ~1~&e^{-t/T_2}&e^{-t/T_2}&e^{-2t/T_2} \\ e^{-t/T_2}&~1~&e^{-2t/T_2}&e^{-t/T_2}\\ e^{-t/T_2}&e^{-2t/T_2}&~1~&e^{-t/T_2}\\ e^{-2t/T_2}&e^{-t/T_2}&e^{-t/T_2}&~1~ \end{bmatrix} .
\end{equation}
To exponentiate the second term, we note that the Hadamard product is with $2\, \MAT P_{03} = 2\, \SIG[2]\otimes\SIG[2]$ and resort briefly to superoperators in order to simplify the exponential as follows:
\begin{equation} \begin{split} &
\MAT{Exp}\big( -\!(2t/T_2)\, \DMAT(\. \KET{\MAT P_{03}} )\, (\MAT P_{30} \otimes \MAT P_{30})\, \big) \\[0.5ex] =\M &
{\sum}_{k=0}^\infty\, \big( -\!(2t/T_2)\, \DMAT(\. \KET{\MAT P_{03}} )\, (\MAT P_{30} \otimes \MAT P_{30}) \,\big)^k \big/\, k! \\[0.5ex] =\M &
\begin{aligned}[t] \MAT P_{00} \otimes \MAT P_{00} ~+~ \DMAT(\. \KET{\MAT P_{30}} ) \sum_{k=1}^\infty\, \frac{{(2t/T_2)}^{2k}}{2k!} ~\cdots & \\
-~ \DMAT( \KET{\MAT P_{03}} )\, & (\MAT P_{30} \otimes \MAT P_{30}) \sum_{k=0}^\infty\, \frac{{(2t/T_2)}^{2k+1}}{(2k+1)!}
\end{aligned} \\ =\M &
\begin{aligned}[t] \MAT P_{00} \otimes \MAT P_{00} ~+~ \DMAT(\. \KET{\MAT P_{30}} )\, (\cosh( 2t / T_2 ) \,-\, 1) ~\cdots & \\
-~ \DMAT(\. \KET{\MAT P_{03}} )\, & (\MAT P_{30} \otimes \MAT P_{30})\,\sinh( 2t / T_2 ) ~.
\end{aligned}
\end{split} \end{equation}
This derivation relies on the facts that $\MAT P_{30} \otimes \MAT P_{30}$ squares to the identity, $\DMAT(\KET{\MAT P_{03}})$ squares to $\DMAT(\KET{\MAT P_{30}})$, and each commutes with the other.

Going back to operator sum notation and abbreviating $\SIG \equiv \SIG(0)$, we thus obtain in all:
\begin{align} &
\SIG(t) \M[.5]=\M[.5] \begin{aligned}[t] \MAT D(t) \odot \big( \SIG \:+\: (\cosh(2t/T_2) - 1)\, \MAT P_{30} \odot \SIG ~\cdots &
\\ -~ \sinh & (2t/T_2)\,\MAT P_{03} \odot (\. \MAT P_{30}\, \SIG\, \MAT P_{30} \.)\. \big)
\end{aligned} \notag \\[1ex] =\M[.5] &
\begin{bmatrix}
\sigma_{00} & \sigma_{01}\, e^{-t/T_2} & \sigma_{02}\, e^{-t/T_2} & \Delta(t;\, \sigma_{03},\, \sigma_{30}) \\[.5ex]
\sigma_{10}\, e^{-t/T_2} & \sigma_{11} & \Delta(t;\, \sigma_{12}, -\sigma_{21}) & \sigma_{13}\, e^{-t/T_2} \\[.5ex]
\sigma_{01}\, e^{-t/T_2} & \Delta(t;\, \sigma_{21},\, -\sigma_{12}) & \sigma_{22} & \sigma_{23}\, e^{-t/T_2} \\[.5ex]
\Delta(t;\, \sigma_{30},\, \sigma_{03}) & \sigma_{31}\, e^{-t/T_2} & \sigma_{32}\, e^{-t/T_2} &  \sigma_{33}
\end{bmatrix} ,
\end{align}
where $\Delta(t;\, x,\, y) \equiv (\cosh(2t/T_2)\. x + \sinh(2t/T_2)\, y))\. \exp(-2t/T_2)$.
From this we see that the anti-diagonal entries decoher into mixtures with their symmetrically placed opposites in the real density matrix.
These mixtures correspond to the real and imaginary parts of the $\rho_{12} = \bar\rho_{21}$ entries in the Hermitian density matrix, otherwise known as \emph{zero-quantum coherences}, which are immune to correlated noise \cite{ErnBodWok:87}.

\section{Epilogue}
We have seen that one can, with some effort, do pretty much everything with the real density matrix that one could with the usual Hermitian one.
This may be useful as a didactic device, or in calculations with experimental (e.g.~NMR) data where it is desirable to keep the experimentally measured values of the observables in sight at all times.
This work is also a good demonstration of the power of Choi matrix decompositions as a means of finding operator sum representations of linear superoperators \cite{Havel!QPT:03}.

Although the Hermitian density matrix is expected to be better suited, by and large, for the purposes of numerical calculations, it is worth emphasizing that for theoretical and/or expository purposes the compact but lucid notation of geometric algebra offers significant advantages over any matrix formalism.
In this regard, we point out that \citet{HavDorFur:03} have recently introduced a \emph{parity-even} (rather than reverse-even, aka Hermitian) multi-qubit density operator via geometric algebra, which generalizes the \emph{multi-particle space-time algebra} introduced for isolated systems to open multi-qubit systems.
It is our hope that in due course such a geometric formulation may provide new insights into some of the conceptual problems that underlie quantum physics.

The existence of the real density matrix is further of some theoretical interest, since it provides a coordinate ring within which one can study the issues of entanglement and decoherence via invariant theoretic methods \cite{GraRotBet:98, Makhlin:02}.
There are intimate connections between invariant theory and geometric algebra, and it is often easier to automate symbolic computations in an invariant ring than it is at the more abstract level of geometric algebra \cite{Sturmfels:93, Havel:97, Havel:01}.

\begin{acknowledgments} \vspace{-0.25\baselineskip}
\noindent The author thanks Nicolas Boulant, David Cory and Chris Doran for useful discussions.
This work was supported by ARO grants DAAD19-01-1-0519, DAAD19-01-1-0678, by DARPA grant MDA972-01-1-0003, and by a grant from the Cambridge-MIT Institute, Ltd.
\end{acknowledgments}

\bibliography{phys,math,csci,self,nmr}
\end{document}
\end

%% file: macros.tex
\makeatletter
\def\@tempa{revtex4}
\newif\ifisRevTeX
\ifx\class@name\@tempa
   \isRevTeXtrue
\else
   \isRevTeXfalse
\fi
\ifisRevTeX
   \renewcommand{\frontmatter@abstractheading}{%
      \begingroup\centering\large\vspace{2\baselineskip}%
      \abstractname\par\endgroup}
\fi
\makeatother

\newcommand{\EMB}{\boldsymbol}

\newcommand{\LAB}[1]{{\mathsf{#1}}}
\newcommand{\FUN}[1]{{\mathrm{#1}}}
\newcommand{\MAT}[1]{{\mathbf{#1}}}

\newcommand{\FLD}[1]{{\mathbb{#1}}}
\newcommand{\ALG}[1]{{\mathcal{#1}}}

\newcommand{\OL}{\overline}

\newcommand{\HALF}{\mathchoice{{\textstyle\frac12}}{\frac12}{\frac12}{\frac12}}

\newcommand{\M}[1][1]{\hspace{#1em}}
\newcommand{\DAB}[1]{\rule[0pt]{0pt}{#1}}
\newcommand{\X}[1][1]{\DAB{#1ex}}

\newcommand{\DMAT}{\MAT{Diag}}

\newcommand{\TR}{\FUN{tr}}

\newcommand{\RHO}{{\EMB\rho}}
\newcommand{\SIG}[1][{}]{\EMB\sigma_{\!#1}}	

\newcommand{\KET}[1]{|\,#1\,\rangle}
\newcommand{\BRA}[1]{\langle\,#1\,|}
\newcommand{\AVG}[1]{\langle\,#1\,\rangle}
\newcommand{\HIP}[2]{\langle\,#1\,|\,#2\,\rangle}

%% file: ms.bbl
\begin{thebibliography}{23}
\expandafter\ifx\csname natexlab\endcsname\relax\def\natexlab#1{#1}\fi
\expandafter\ifx\csname bibnamefont\endcsname\relax
  \def\bibnamefont#1{#1}\fi
\expandafter\ifx\csname bibfnamefont\endcsname\relax
  \def\bibfnamefont#1{#1}\fi
\expandafter\ifx\csname citenamefont\endcsname\relax
  \def\citenamefont#1{#1}\fi
\expandafter\ifx\csname url\endcsname\relax
  \def\url#1{\texttt{#1}}\fi
\expandafter\ifx\csname urlprefix\endcsname\relax\def\urlprefix{URL }\fi
\providecommand{\bibinfo}[2]{#2}
\providecommand{\eprint}[2][]{\url{#2}}

\bibitem[{\citenamefont{Bloch}(1946)}]{Bloch:46}
\bibinfo{author}{\bibfnamefont{F.}~\bibnamefont{Bloch}},
  \bibinfo{journal}{Phys. Rev.} \textbf{\bibinfo{volume}{70}},
  \bibinfo{pages}{460} (\bibinfo{year}{1946}).

\bibitem[{\citenamefont{Feynman et~al.}(1957)\citenamefont{Feynman, Vernon, and
  Hellwarth}}]{FeyVerHel:57}
\bibinfo{author}{\bibfnamefont{R.~P.} \bibnamefont{Feynman}},
  \bibinfo{author}{\bibfnamefont{F.~L.} \bibnamefont{Vernon}},
  \bibnamefont{and} \bibinfo{author}{\bibfnamefont{R.~W.}
  \bibnamefont{Hellwarth}}, \bibinfo{journal}{J. Appl. Phys.}
  \textbf{\bibinfo{volume}{28}}, \bibinfo{pages}{49} (\bibinfo{year}{1957}).

\bibitem[{\citenamefont{Ernst et~al.}(1987)\citenamefont{Ernst, Bodenhausen,
  and Wokaun}}]{ErnBodWok:87}
\bibinfo{author}{\bibfnamefont{R.~R.} \bibnamefont{Ernst}},
  \bibinfo{author}{\bibfnamefont{G.}~\bibnamefont{Bodenhausen}},
  \bibnamefont{and} \bibinfo{author}{\bibfnamefont{A.}~\bibnamefont{Wokaun}},
  \emph{\bibinfo{title}{Principles of Nuclear Magnetic Resonance in One and Two
  Dimensions}} (\bibinfo{publisher}{Oxford Univ. Press},
  \bibinfo{address}{U.K.}, \bibinfo{year}{1987}).

\bibitem[{\citenamefont{Mahler and Weberruss}(1998)}]{MahleWeber:98}
\bibinfo{author}{\bibfnamefont{G.}~\bibnamefont{Mahler}} \bibnamefont{and}
  \bibinfo{author}{\bibfnamefont{V.~A.} \bibnamefont{Weberruss}},
  \emph{\bibinfo{title}{Quantum Networks (2nd ed.)}}
  (\bibinfo{publisher}{Springer-Verlag}, \bibinfo{address}{Berlin, Heidelberg,
  New York}, \bibinfo{year}{1998}).

\bibitem[{\citenamefont{Havel and Doran}(2002{\natexlab{a}})}]{HavelDoran:02}
\bibinfo{author}{\bibfnamefont{T.~F.} \bibnamefont{Havel}} \bibnamefont{and}
  \bibinfo{author}{\bibfnamefont{C.}~\bibnamefont{Doran}}, in
  \emph{\bibinfo{booktitle}{Quantum Information and Computation}}, edited by
  \bibinfo{editor}{\bibfnamefont{S.~J.} \bibnamefont{{Lomonaco, Jr.}}}
  \bibnamefont{and} \bibinfo{editor}{\bibfnamefont{H.~E.} \bibnamefont{Brandt}}
  (\bibinfo{publisher}{Am. Math. Soc., Providence, RI},
  \bibinfo{year}{2002}{\natexlab{a}}), vol. \bibinfo{volume}{305} of
  \emph{\bibinfo{series}{Contempoary Mathematics}}, pp.
  \bibinfo{pages}{81--100}, \bibinfo{note}{(see also LANL preprint
  \texttt{quant-ph/0004031}).}

\bibitem[{\citenamefont{Doran and Lasenby}(2003)}]{DoranLasen:03}
\bibinfo{author}{\bibfnamefont{C.}~\bibnamefont{Doran}} \bibnamefont{and}
  \bibinfo{author}{\bibfnamefont{A.}~\bibnamefont{Lasenby}},
  \emph{\bibinfo{title}{Geometric Algebra for Physicists}}
  (\bibinfo{publisher}{Cambridge Univ. Press}, \bibinfo{address}{Cambridge,
  U.K.}, \bibinfo{year}{2003}).

\bibitem[{\citenamefont{Hestenes}(2003)}]{Hestenes:03}
\bibinfo{author}{\bibfnamefont{D.}~\bibnamefont{Hestenes}},
  \bibinfo{journal}{Am. J. Phys.} \textbf{\bibinfo{volume}{71}},
  \bibinfo{pages}{104} (\bibinfo{year}{2003}).

\bibitem[{\citenamefont{Baylis}(1999)}]{Baylis:99}
\bibinfo{author}{\bibfnamefont{W.~E.} \bibnamefont{Baylis}},
  \emph{\bibinfo{title}{Electrodynamics: {A} Modern Geometric Approach}}
  (\bibinfo{publisher}{{Birkh\"auser} Pub.~Co.}, \bibinfo{year}{1999}).

\bibitem[{\citenamefont{Havel}(2003)}]{Havel!QPT:03}
\bibinfo{author}{\bibfnamefont{T.~F.} \bibnamefont{Havel}},
  \bibinfo{journal}{J. Math. Phys.} \textbf{\bibinfo{volume}{44}},
  \bibinfo{pages}{534} (\bibinfo{year}{2003}).

\bibitem[{\citenamefont{Havel et~al.}(2001)\citenamefont{Havel, Sharf, Viola,
  and Cory}}]{HaShViCo:01}
\bibinfo{author}{\bibfnamefont{T.~F.} \bibnamefont{Havel}},
  \bibinfo{author}{\bibfnamefont{Y.}~\bibnamefont{Sharf}},
  \bibinfo{author}{\bibfnamefont{L.}~\bibnamefont{Viola}}, \bibnamefont{and}
  \bibinfo{author}{\bibfnamefont{D.~G.} \bibnamefont{Cory}},
  \bibinfo{journal}{Phys. Lett. A} \textbf{\bibinfo{volume}{280}},
  \bibinfo{pages}{282} (\bibinfo{year}{2001}).

\bibitem[{\citenamefont{Nielsen and Chuang}(2000)}]{NielsChuan:00}
\bibinfo{author}{\bibfnamefont{M.~A.} \bibnamefont{Nielsen}} \bibnamefont{and}
  \bibinfo{author}{\bibfnamefont{I.~L.} \bibnamefont{Chuang}},
  \emph{\bibinfo{title}{Quantum Computation and Quantum Information}}
  (\bibinfo{publisher}{Cambridge Univ. Press}, \bibinfo{year}{2000}).

\bibitem[{\citenamefont{Havel et~al.}(2002)\citenamefont{Havel, Cory, Lloyd,
  Fortunato, Pravia, Teklemariam, Weinstein, Bhattacharyya, and
  Hou}}]{HCLBFPTWBH:02}
\bibinfo{author}{\bibfnamefont{T.~F.} \bibnamefont{Havel}},
  \bibinfo{author}{\bibfnamefont{D.~G.} \bibnamefont{Cory}},
  \bibinfo{author}{\bibfnamefont{S.}~\bibnamefont{Lloyd}},
  \bibinfo{author}{\bibfnamefont{N.~B. E.~M.} \bibnamefont{Fortunato}},
  \bibinfo{author}{\bibfnamefont{M.~A.} \bibnamefont{Pravia}},
  \bibinfo{author}{\bibfnamefont{G.}~\bibnamefont{Teklemariam}},
  \bibinfo{author}{\bibfnamefont{Y.~S.} \bibnamefont{Weinstein}},
  \bibinfo{author}{\bibfnamefont{A.}~\bibnamefont{Bhattacharyya}},
  \bibnamefont{and} \bibinfo{author}{\bibfnamefont{J.}~\bibnamefont{Hou}},
  \bibinfo{journal}{Am. J. Phys.} \textbf{\bibinfo{volume}{70}},
  \bibinfo{pages}{345} (\bibinfo{year}{2002}).

\bibitem[{\citenamefont{Pittenger and Rubin}(2000)}]{PitteRubin:00a}
\bibinfo{author}{\bibfnamefont{A.~O.} \bibnamefont{Pittenger}}
  \bibnamefont{and} \bibinfo{author}{\bibfnamefont{M.~H.} \bibnamefont{Rubin}},
  \bibinfo{journal}{Phys. Rev. A} \textbf{\bibinfo{volume}{62}},
  \bibinfo{pages}{032313} (\bibinfo{year}{2000}).

\bibitem[{\citenamefont{{L\"utkepohl}}(1996)}]{Lutkepohl:96}
\bibinfo{author}{\bibfnamefont{H.}~\bibnamefont{{L\"utkepohl}}},
  \emph{\bibinfo{title}{Handbook of Matrices}} (\bibinfo{publisher}{John Wiley
  \& Sons}, \bibinfo{address}{New York, NY}, \bibinfo{year}{1996}).

\bibitem[{\citenamefont{Somaroo et~al.}(1998)\citenamefont{Somaroo, Cory, and
  Havel}}]{SomCorHav:98}
\bibinfo{author}{\bibfnamefont{S.~S.} \bibnamefont{Somaroo}},
  \bibinfo{author}{\bibfnamefont{D.~G.} \bibnamefont{Cory}}, \bibnamefont{and}
  \bibinfo{author}{\bibfnamefont{T.~F.} \bibnamefont{Havel}},
  \bibinfo{journal}{Phys. Lett. A} \textbf{\bibinfo{volume}{240}},
  \bibinfo{pages}{1} (\bibinfo{year}{1998}).

\bibitem[{\citenamefont{Najfeld and Havel}(1995)}]{NajfeHavel:95a}
\bibinfo{author}{\bibfnamefont{I.}~\bibnamefont{Najfeld}} \bibnamefont{and}
  \bibinfo{author}{\bibfnamefont{T.~F.} \bibnamefont{Havel}},
  \bibinfo{journal}{Adv. Appl. Math.} \textbf{\bibinfo{volume}{16}},
  \bibinfo{pages}{321} (\bibinfo{year}{1995}).

\bibitem[{\citenamefont{Havel and Doran}(2002{\natexlab{b}})}]{HaDo$AGACSE:02}
\bibinfo{author}{\bibfnamefont{T.~F.} \bibnamefont{Havel}} \bibnamefont{and}
  \bibinfo{author}{\bibfnamefont{C.~J.~L.} \bibnamefont{Doran}}, in
  \emph{\bibinfo{booktitle}{Applications of Geometric Algebra in Science and
  Engineering}}, edited by
  \bibinfo{editor}{\bibfnamefont{L.}~\bibnamefont{Dorst}},
  \bibinfo{editor}{\bibfnamefont{C.}~\bibnamefont{Doran}}, \bibnamefont{and}
  \bibinfo{editor}{\bibfnamefont{J.}~\bibnamefont{Lasenby}}
  (\bibinfo{publisher}{{Birkh\"auser}, Boston, MA},
  \bibinfo{year}{2002}{\natexlab{b}}).

\bibitem[{\citenamefont{Havel et~al.}(2003)\citenamefont{Havel, Doran, and
  Furuta}}]{HavDorFur:03}
\bibinfo{author}{\bibfnamefont{T.~F.} \bibnamefont{Havel}},
  \bibinfo{author}{\bibfnamefont{C.~J.~L.} \bibnamefont{Doran}},
  \bibnamefont{and} \bibinfo{author}{\bibfnamefont{S.}~\bibnamefont{Furuta}},
  \bibinfo{journal}{Proc. R. Soc. Lond. A}  (\bibinfo{year}{2003}),
  \bibinfo{note}{in press.}

\bibitem[{\citenamefont{Grassl et~al.}(1998)\citenamefont{Grassl, {R\"otteler},
  and Beth}}]{GraRotBet:98}
\bibinfo{author}{\bibfnamefont{M.}~\bibnamefont{Grassl}},
  \bibinfo{author}{\bibfnamefont{M.}~\bibnamefont{{R\"otteler}}},
  \bibnamefont{and} \bibinfo{author}{\bibfnamefont{T.}~\bibnamefont{Beth}},
  \bibinfo{journal}{Phys. Rev. A} \textbf{\bibinfo{volume}{58}},
  \bibinfo{pages}{1833} (\bibinfo{year}{1998}).

\bibitem[{\citenamefont{Makhlin}(2002)}]{Makhlin:02}
\bibinfo{author}{\bibfnamefont{Y.}~\bibnamefont{Makhlin}},
  \bibinfo{journal}{Quantum Inform. Processing} \textbf{\bibinfo{volume}{1}},
  \bibinfo{pages}{243} (\bibinfo{year}{2002}).

\bibitem[{\citenamefont{Havel}(1997)}]{Havel:97}
\bibinfo{author}{\bibfnamefont{T.~F.} \bibnamefont{Havel}}, in
  \emph{\bibinfo{booktitle}{Automated Deduction in Geometry}}, edited by
  \bibinfo{editor}{\bibfnamefont{D.}~\bibnamefont{Wang}}
  (\bibinfo{publisher}{Springer-Verlag}, \bibinfo{year}{1997}), vol.
  \bibinfo{volume}{1360} of \emph{\bibinfo{series}{Lect. Notes in Artif.
  Intellig.}}, pp. \bibinfo{pages}{102--114}.

\bibitem[{\citenamefont{Havel}(2001)}]{Havel:01}
\bibinfo{author}{\bibfnamefont{T.~F.} \bibnamefont{Havel}}, in
  \emph{\bibinfo{booktitle}{Automated Deduction in Geometry}}, edited by
  \bibinfo{editor}{\bibfnamefont{J.}~\bibnamefont{Richter-Gebert}}
  \bibnamefont{and} \bibinfo{editor}{\bibfnamefont{D.}~\bibnamefont{Wang}}
  (\bibinfo{publisher}{Springer-Verlag}, \bibinfo{year}{2001}), vol.
  \bibinfo{volume}{2061} of \emph{\bibinfo{series}{Lect. Notes in Artif.
  Intellig.}}, pp. \bibinfo{pages}{228--245}.

\bibitem[{\citenamefont{Sturmfels}(1993)}]{Sturmfels:93}
\bibinfo{author}{\bibfnamefont{B.}~\bibnamefont{Sturmfels}},
  \emph{\bibinfo{title}{Algorithms in Invariant Theory}}
  (\bibinfo{publisher}{Springer-Verlag, Wien, New York}, \bibinfo{year}{1993}).

\end{thebibliography}
